\documentclass[11pt]{article}
\usepackage{setspace}
\usepackage[margin=1in]{geometry}
\usepackage{amsmath,amsthm,amssymb}
\usepackage{bm,bbm}
\usepackage{color}
\usepackage{tikz}
\usepackage{subcaption}
\usepackage{multirow}
\allowdisplaybreaks

\newtheorem{lemma}{Lemma}[section]

\newtheorem{theorem}[lemma]{Theorem}
\newtheorem{corollary}[lemma]{Corollary}

\newtheorem{example1}[lemma]{Example}
\newtheorem{rem1}[lemma]{Remark}
\newtheorem{assumption}[lemma]{Assumption}
\newtheorem{alg1}[lemma]{Algorithm}
\newtheorem{me1}[lemma]{Mechanism}
\newenvironment{remark}{\begin{rem1}\rm}{\end{rem1}}
\newenvironment{example}{\begin{example1}\rm}{\end{example1}}

\newcommand{\bbr}{\mathbb{R}}
\renewcommand{\P}{\mathbb{P}}
\newcommand{\T}{\top}
\newcommand{\diag}{\operatorname{diag}}
\newcommand{\ind}{\mathbbm{1}}
\renewcommand{\d}{\partial}

\newcommand{\dG}{\frac{\d}{\d \Gamma}}

\begin{document}

\title{Capital Regulation under Price Impacts and Dynamic Financial Contagion}
\author{Zachary Feinstein\footnote{Stevens Institute of Technology, School of Business, Hoboken, NJ 07030, USA, {\tt zfeinste@stevens.edu}.}\\[0.7ex] \textit{Stevens Institute of Technology}}
\date{\today}
\maketitle

\begin{abstract}
We construct a continuous time model for price-mediated contagion precipitated by a common exogenous stress to the banking book of all firms in the financial system.  In this setting, firms are constrained so as to satisfy a risk-weight based capital ratio requirement.  We use this model to find analytical bounds on the risk-weights for assets as a function of the market liquidity.  Under these appropriate risk-weights, we find existence and uniqueness for the joint system of firm behavior and the asset prices.  We further consider an analytical bound on the firm liquidations, which allows us to construct exact formulas for stress testing the financial system with deterministic or random stresses.  
Numerical case studies are provided to demonstrate various implications of this model and analytical bounds.
\end{abstract}\vspace{0.2cm}

\section{Introduction}\label{sec:intro}
Financial contagion occurs when the \emph{negative} actions of one bank or firm causes the distress of a separate bank or firm.  Such events are of critical importance due to their relation to systemic risk.  In this work we consider price-mediated contagion that occurs through impacts to mark-to-market wealth as firms hold overlapping portfolios.  Price-mediated contagion can occur due to the price impacts of liquidations in a crisis and can be exacerbated by pro-cyclical regulations.  Importantly, this kind of contagion can be self-reinforcing, causing extreme events and ultimately a systemic crisis as witnessed in, e.g., the 2007-2009 financial crisis.

Systemic risk and financial contagion has been studied in a network of interbank payments by \cite{EN01}.  We refer to \cite{AW_15} for a review of this payment network model and extensions thereof to include, e.g., bankruptcy costs.  
The focus of this paper is on price-mediated contagion and fire sales.  This single contagion channel causes impacts globally to all other firms due to mark-to-market accounting.  As prices drop due to the liquidations of one bank, the value of the assets of all other banks are also impacted.  
The model from \cite{EN01} has been extended to consider fire sales and price-mediated contagion in a static, one-period, system by works such as \cite{CFS05,GK10,CLY14,AW_15,AFM16,feinstein2015illiquid,feinstein2016leverage,feinstein2018currency,bichuch2018loans}.  
Price-mediated contagion and fire sales have been studied in other works without the inclusion of interbank payment networks. 
This has been undertaken in a static setting by \cite{greenwood2015,CS16,braouezec2016risk,braouezec2017strategic}, in a discrete time setting in \cite{caccioli2014,CL15}, and in continuous time by \cite{CW13,CW14}.

In this work we will be extending the model of \cite{braouezec2016risk,braouezec2017strategic} to incorporate true time dynamics.  Those works present a static price-mediated contagion due to deleveraging and the need to satisfy a capital ratio requirement.  In particular, we will focus on the case in which firms liquidate assets during a crisis due to risk-weighted capital requirement constraints.  These capital requirements will be described by the ratio of equity over risk-weighted assets.  We focus on those works as they include methodology for calibrating the model to public data, but also include equilibrium liquidations and prices that in reality occur over time.  Herein we will consider a continuous time model for these equilibrium liquidations and price movements.  We will demonstrate that such a model has useful mathematical properties, notably uniqueness of the clearing prices in time.  This is in contrast to the static models of, e.g., \cite{CFS05,feinstein2016leverage,braouezec2017strategic} in which fire sales due to capital ratios can result in multiple equilibria.  Further, by incorporating time dynamics, we are able to consider the first-mover advantage in which the first firm to engage in the fire sale will receive a higher price than later firms.  This is not accounted for in any of the static models discussed previously.

Briefly, the risk-weighted capital ratio that we consider in this work is featured in, e.g., the Basel Accords and is defined by a firm's capital divided by its risk-weighted assets. For more details, we refer to \cite{braouezec2016risk,braouezec2017strategic}.  Officially, in Basel III, the total capital is defined as the sum of Tier 1 and Tier 2 capital.  In this work, we do not consider a distinction between different types of capital.  The risk-weighted assets are defined as being a weighted sum of the mark-to-market assets.  Conceptually, the riskier an asset the greater its risk-weight.  
The risk-weights of credit portfolios are given by, e.g., the Basel Accords or national laws, and often determined by internal models of each institution. 
Basel regulations state that the risk-based capital ratio must never be below 8\%.  When a firm is constrained by this ratio, the firm will typically need to liquidate assets in order to reduce liabilities as issuing equity in such a scenario is often untenable or excessively costly \cite{hanson2011,greenlaw2012}.  However, these liquidations can and will cause price impacts on the risk-weighted assets.  This causes feedback effects which causes that same firm to liquidate further assets as well as negatively effects the risk-weighted capital ratio of all other firms.

As stated, the static model was studied in, e.g., \cite{CFS05,braouezec2016risk,braouezec2017strategic,feinstein2016leverage}.  In those works, uniqueness of the prices and liquidations cannot be guaranteed for most financial systems; this is true in settings with fire sales only (i.e., without interbank assets and liabilities).  Further, if the price impact is too large it is found that banks can no longer satisfy their capital ratio requirement even if they hold only tradable assets.  In contrast, we will demonstrate that, in this special setting and in continuous time, a firm will never need to sell all assets, though may asymptote its asset holdings to 0.  Of course, as expected, when firms hold liquid or untradable assets whose value fluctuates over time, this model would allow for firms to liquidate all tradeable assets and become insolvent.  In fact, we will relate the price impacts of the illiquid assets to appropriate risk-weights.  We will also demonstrate that if the risk-weight were set too low in relation to price impacts, the firm will be forced to purchase assets to drive up the price rather than liquidate.

The primary goal of this paper is to model the behavior of banks so that they satisfy this capital ratio requirement continuously in time under price impacts.  The use of a dynamic model is important as the Basel regulations enforce the risk-weighted capital ratio to exceed the threshold at all times.  In particular, we will consider the situation in which multiple banks may be at the regulatory threshold and behaving in the required manner so as to consider the implications of financial contagion to systemic risk.  In utilizing the proposed model, we will consider appropriate choices for the risk-weights as a function of market liquidity.
Additionally, in proposing the continuous time model for bank behavior, we find an analytical bound to the firm behavior.  This is particularly of value as it allows us to consider a distribution of outcomes for the health of the system directly under randomized stress tests.

This paper is organized as follows.  Section~\ref{sec:regulatory} presents the risk-weighted capital ratio. 
Section~\ref{sec:1-bank} proposes the differential model for the actions of a single bank system with a single, representative, tradable illiquid asset.  
This is extended in Section~\ref{sec:n-bank} to provide existence and uniqueness results in a $n$ bank financial system.  
The modeling is completed in Section~\ref{sec:m-asset} to present a market with a $n$ bank financial system and $m$ tradable illiquid assets.  As this model has no closed-form solution in general, we propose an analytical approximation that bounds the system response for stress testing purposes in Section~\ref{sec:stresstest}.  These analytical results allow for a bound on, e.g., the probability that the terminal asset price is above some threshold in a probabilistic setting.  Numerical case studies are provided in Section~\ref{sec:casestudy} to demonstrate simple insights from this model and provide numerical accuracy of the analytical bounds from Section~\ref{sec:stresstest}.  The proofs are presented in the appendix.

\section{The Risk-Weighted Capital Ratio}\label{sec:regulatory}

\begin{figure}[h!]
\centering
\begin{tikzpicture}
\draw[draw=none] (0,6.5) rectangle (6,7) node[pos=.5]{\bf Initial Banking Book};
\draw[draw=none] (0,6) rectangle (3,6.5) node[pos=.5]{\bf Assets};
\draw[draw=none] (3,6) rectangle (6,6.5) node[pos=.5]{\bf Liabilities};

\filldraw[fill=blue!20!white,draw=black] (0,5) rectangle (3,6) node[pos=.5,style={align=center}]{Liquid \\ $x$};
\filldraw[fill=yellow!20!white,draw=black] (0,1.8) rectangle (3,5) node[pos=.5,style={align=center}]{Illiquid \\ (Tradable) \\ $s$};
\filldraw[fill=green!20!white,draw=black] (0,0) rectangle (3,1.8) node[pos=.5,style={align=center}]{Illiquid \\ (Nontradable) \\ $\ell$};

\filldraw[fill=purple!20!white,draw=black] (3,3) rectangle (6,6) node[pos=.5,style={align=center}]{Total \\ $\bar p$};
\filldraw[fill=orange!20!white,draw=black] (3,0) rectangle (6,3) node[pos=.5,style={align=center}]{Capital \\ $x + s + \ell - \bar p$};

\draw[->,line width=1mm] (6.5,3) -- (8.5,3);

\draw[draw=none] (9,6.5) rectangle (15,7) node[pos=.5]{\bf Updated Banking Book};
\draw[draw=none] (9,6) rectangle (12,6.5) node[pos=.5]{\bf Assets};
\draw[draw=none] (12,6) rectangle (15,6.5) node[pos=.5]{\bf Liabilities};

\filldraw[fill=blue!20!white,draw=none] (9,5) rectangle (12,6) node[pos=.5,style={align=center}]{Liquid \\ $x$};
\filldraw[fill=blue!20!white,draw=none,opacity=0.5] (9,4.5) rectangle (12,5) node[pos=.5,style={align=center}]{$\Psi(t)$};
\filldraw[fill=yellow!20!white,draw=none,opacity=0.5] (9,4.5) rectangle (12,5) node[pos=.5,style={align=center}]{$\Psi(t)$};
\draw[draw=none] (9,4.5) rectangle (12,5) node[pos=.5]{$\Psi(t)$};
\draw[dotted] (9,5) -- (12,5);
\filldraw[fill=yellow!20!white,draw=black] (9,2.8) rectangle (12,4.5) node[pos=.5,style={align=center}]{Illiquid \\ (Tradable) \\ $(s-\Gamma(t))q(t)$};
\filldraw[fill=green!20!white,draw=black] (9,1) rectangle (12,2.8) node[pos=.5,style={align=center}]{Illiquid \\ (Nontradable) \\ $\ell$};
\filldraw[fill=yellow!20!white,draw=black] (9,0) rectangle (12,1);

\filldraw[fill=purple!20!white,draw=black] (12,3) rectangle (15,6) node[pos=.5,style={align=center}]{Total \\ $\bar p$};
\filldraw[fill=orange!20!white,draw=black] (12,1) rectangle (15,3) node[pos=.5,style={align=center}]{Capital \\ $x + \Psi(t) + $ \\ $(s-\Gamma(t))q(t)$ \\ $+ \ell - \bar p$};
\filldraw[fill=orange!20!white,draw=black] (12,0) rectangle (15,1);
\draw (9,0) rectangle (15,6);
\draw (12,0) -- (12,6);

\begin{scope}
    \clip (9,0) rectangle (15,1);
    \foreach \x in {-9,-8.5,...,15}
    {
        \draw[line width=.5mm] (9+\x,0) -- (15+\x,6);
    }
\end{scope}
\end{tikzpicture}
\caption{Stylized banking book for a firm before and after price and liquidation updates with 1 tradable illiquid asset.}
\label{fig:balance_sheet}
\end{figure}
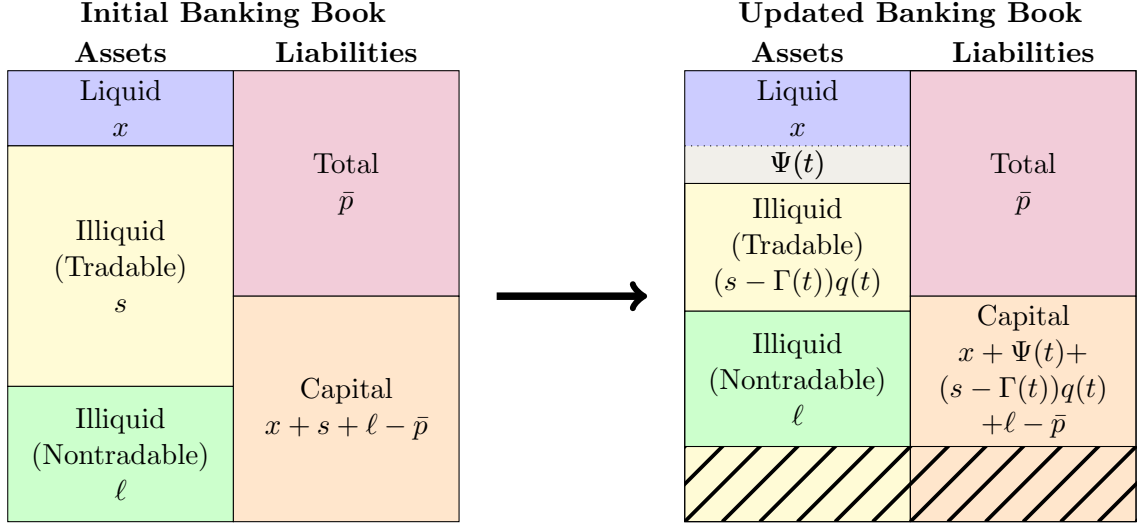 

Consider a firm with stylized banking book depicted in Figure~\ref{fig:balance_sheet}, but with $m \geq 1$ tradable illiquid assets.  That is, at time $0$, the firm has assets split between liquid investments (e.g., cash or otherwise zero risk-weighted assets) denoted by $x \geq 0$, tradable illiquid investments (e.g., tradable credit positions) denoted by $s \in \bbr^m_+$, and nontradable illiquid investments (e.g., residential loans) denoted by $\ell \geq 0$.  For simplicity and without loss of generality, we will assume that the initial price of all assets is $1$, thus the mark-to-market assets for the bank at time $0$ is equal to $x + s + \ell$.  The firm has liabilities in the total amount of $\bar p \geq 0$.  For simplicity in this work, we will assume that all liabilities are not held by any other firms in this system; additionally, we will assume that no liabilities come due during the (short time horizon) of the fire sale cascade under study, but are liquid enough that they can frictionlessly be paid off early with liquid assets.  The capital of the firm, at time $0$, is thus provided by $x + s + \ell - \bar p$.  For further simplicity, in Sections~\ref{sec:1-bank} and~\ref{sec:n-bank} we will assume a single, representative, tradable illiquid asset only.  This is along the lines of the modeling undertaken in, e.g., \cite{CFS05,AFM16,braouezec2017strategic}.

Capital ratios are used for regulatory purposes to bound the risk of financial institutions.  We will assume that the tradable illiquid assets have risk-weight $\alpha \in \bbr^m_+$ and price process $q: [0,T] \to \bbr_{++}^m$ (with $q_k(0) = 1$ for every asset $k$).
The bank may liquidate assets over time.  We will assume that, at time $t$, they liquidate the tradable illiquid assets at a rate of $\gamma(t) \in \bbr^m$.  The total amount of cash gained from liquidations up to time $t$ is provided by $\Psi(t) = \int_0^t \gamma(u)^\T q(u)du \in \bbr$ and the total number of units liquidated up to time $t$ is provided by $\Gamma(t) = \int_0^t \gamma(u)du \in \bbr^m$.  Thus, as depicted in Figure~\ref{fig:balance_sheet}, at time $t$, the liquid assets for the firm are provided by $x + \Psi(t)$ and the tradable illiquid assets by $(s-\Gamma(t))^\T q(t)$.  Throughout this work we will assume that prices drop over time and as a function of the liquidations, so the total assets and therefore also capital will drop over time as shown by the crossed out portions of the banking book in Figure~\ref{fig:balance_sheet}.  More discussion on the price changes will be provided in Section~\ref{sec:1-bank} below.  Additionally, the liquid assets have 0 risk weight ($\alpha_x = 0$) and nontradable illiquid assets have risk-weight $\alpha_{\ell} \geq 0$.  In settings with more than one bank, we allow for the risk-weights of the nontradable assets to be heterogeneous between institutions.

The capital ratio for a firm at time $t$ is given by total capital divided by the risk-weighted assets.  Mathematically, this is formulated as
\begin{align}
\label{eq:cr} \theta(t) &= \frac{\left(x + \Psi(t) + \sum_{k = 1}^m \left[s_k - \Gamma_k(t)\right] q_k(t) + \ell - \bar p\right)^+}{\sum_{k = 1}^m \alpha_k \left[s_k - \Gamma_k(t)\right] q_k(t) + \alpha_{\ell} \ell}.
\end{align}
The capital ratio requirement specifies that all institutions must satisfy the condition that $\theta(t) \geq \theta_{\min}$ for all times $t$ for some minimal threshold $\theta_{\min} > 0$.
We wish to note that the capital ratio is related to the leverage ratio (assets over equity) by choosing $\alpha_k = 1$ for every tradable asset $k$, $\alpha_{\ell} = 1$, and $\theta_{\min} = 1/\lambda_{\max}$ for leverage requirement $\lambda_{\max} > 0$.  This relationship is utilized in Example~\ref{ex:leverage}.

\begin{assumption}\label{ass:risk-weight}
Throughout this work, we assume $\alpha_k,\theta_{\min} > 0$ with $\alpha_k\theta_{\min} < 1$ for all assets $k$ and $\alpha_\ell \geq 0$.  Additionally, any firm in the financial system will be assumed to satisfy the capital ratio at the initial time $0$, i.e., $\theta(0) \geq \theta_{\min}$.
\end{assumption}
\begin{remark}\label{rem:alphatheta}
If $\alpha_{k^*} \theta_{\min} \geq 1$ for some asset $k^*$, the assumption that the capital ratio at time $0$ is above the regulatory threshold guarantees that the risk-weighted capital ratio is nonincreasing in the price of that asset.  To see this we note that, at time $t = 0$ (i.e., before any intervention from the bank):
\[\frac{(x + \sum_{k = 1}^m s_k q_k(0) + \ell - \bar p)^+}{\sum_{k = 1}^m \alpha_k s_k q_k(0) + \alpha_{\ell} \ell} \geq \theta_{\min} \; \Leftrightarrow \; x + (1 - \alpha_{\ell} \theta_{\min})\ell - \bar p \geq \sum_{k = 1}^m (\alpha_k \theta_{\min} - 1) s_k q_k(0).\]
However, the capital ratio being nonincreasing in the price of asset $k^*$ is contrary to the understanding of how a regulatory threshold usually works.  In particular, for the considerations of this paper, this monotonicity implies that, as the price drops in that asset (without the intervention of the firm), the bank will always satisfy the capital regulation, and thus no rebalancing of assets will ever need to occur.

The assumption that $\alpha_{k} > 0$ for all assets $k$ is slightly stronger than exhibited in reality, as there are liquid assets that are not cash-like instruments.  However, as found in the main results below (see, e.g., Theorem~\ref{thm:unique}), if an asset has risk-weight of 0 then it must have no market impacts from liquidation.  Thus, from the modeling perspective of this work, liquid assets exhibit many behaviors of cash and are merged for the purposes of this work.
\end{remark}

\section{Continuous Time Capital Ratio Requirements}\label{sec:model}

\subsection{Capital Ratio Requirements for a Single Bank System with Single Representative Tradable Illiquid Asset}\label{sec:1-bank}
In this section we consider a single firm attempting to satisfy its risk-weighted capital ratio when subject to price impacts.  We will consider this in continuous time and determine conditions that provide unique liquidations for the bank to satisfy the capital requirement.  In particular, we determine a condition relating the risk-weight and the price impacts.

Consider a single bank with a single tradable illiquid asset.  As the crisis we wish to model is generically on a short time horizon, we will consider all price impacts to be permanent for the duration of the considered time $[0,T] \subseteq \bbr_+$.  Further, we will assume the price of the illiquid asset is subject to market impacts given by a nonincreasing inverse demand function $F: \bbr_+ \times \bbr \to \bbr_{++}$ such that $F(0,0) = 1$.  That is, $F(t,\Gamma)$ is a function of time and units sold; the inclusion of time allows for exogenous shocks, e.g., $F(t,\Gamma) = \exp(-at\ind_{\{t < T\}} - aT\ind_{\{t \geq T\}}) f_\Gamma(\Gamma)$ for some inverse demand function $f_\Gamma: \bbr \to \bbr_{++}$.  For mathematical simplicity we will restrict ourselves to the situation in which we can decouple the exogenous effects from time and the endogenous effects from firm behavior. Through the inverse demand function we find the price of the asset $q(t) = F(t,\Gamma(t))$.
\begin{assumption}\label{ass:idf}
Throughout this work we assume that $F(t,\Gamma) = f_t(t)f_\Gamma(\Gamma)$ for continuously differentiable and nonincreasing function $f_t: \bbr_+ \to (0,1]$ and twice continuously differentiable and nonincreasing function $f_\Gamma: \bbr \to \bbr_{++}$ where $f_t(0) = f_\Gamma(0) = 1$.
\end{assumption}
\begin{remark}\label{rem:exponential}
Throughout this paper we assume Assumption~\ref{ass:idf}, i.e., the clearing prices follow the path $f_t(t)f_\Gamma(\Gamma(t))$.  However, realistically the price effects from time occur due to asset liquidations outside of the firm of interest, i.e., $F(t,\Gamma) = f_\Gamma(\eta(t) + \Gamma)$ for some (nondecreasing) exogenous liquidation function $\eta$.  Letting the full inverse demand function be defined by the exponential inverse demand function with strictly positive price impact, i.e., $f_\Gamma(\Gamma) := \exp(-b\Gamma)$ with $b > 0$, then $F(t,\Gamma) = \exp(-b\eta(t))\exp(-b\Gamma)$.  In particular, we can provide the one-to-one correspondence: $f_t(t) = \exp(-b\eta(t))$ and $\eta(t) = -\frac{1}{b}\log(f_t(t))$.  If $f_\Gamma$ were chosen otherwise, the exogenous liquidations would need to be defined as a function of bank liquidations $\Gamma$ as well.
\end{remark}

Recall the setting described in Section~\ref{sec:regulatory}.  That is, consider the firm with initial banking book made up of liquid assets of $x \geq 0$, liabilities of $\bar p \geq 0$, and illiquid holdings of $s,\ell \geq 0$ at time $0$ with a no-short selling constraint.  The capital ratio is given by \eqref{eq:cr}.
As mentioned previously, we will assume that $\theta(0) \geq \theta_{\min} > 0$ so that the firm satisfies the capital ratio requirement at time $0$.  The change in $\theta$ over time when $\theta(t) > 0$ (i.e., with positive capital) is thus given by:
\begin{align}
\label{eq:dot-theta} \dot \theta(t) &= \frac{\dot q(t) [s - \Gamma(t)] [\alpha(\bar p - x - \Psi(t) - \ell) + \alpha_{\ell}\ell] + \alpha\dot \Gamma(t) q(t) ([s - \Gamma(t)]q(t) - [\bar p - x - \Psi(t) - \ell])}{(\alpha [s - \Gamma(t)]q(t) + \alpha_{\ell}\ell)^2} 
\end{align}
where the change in prices and recovered cash from liquidations are provided by
\begin{align}
\label{eq:dot-q} \dot q(t) &= f_t'(t)f_\Gamma(\Gamma(t)) + \dot \Gamma(t) f_t(t)f_\Gamma'(\Gamma(t)),\\
\label{eq:dot-psi} \dot \Psi(t) &= \dot \Gamma(t) q(t).
\end{align}
As a simplifying assumption, no liquidations will occur except if $\theta(t) \leq \theta_{\min}$.  Therefore the first time that the firm takes actions is at time $\tau$ such that $f_t(\tau) = \bar q := \frac{\bar p - x - (1-\alpha_{\ell}\theta_{\min})\ell}{(1-\alpha\theta_{\min}) s}$.  If $\inf_{t \in [0,T]} f_t(t) > \bar q$ then no fire sale will occurs.  Once the firm starts acting, we assume that it does so only to the extent that it remains at the capital ratio requirement.  Assuming it is possible (as proven later in this section) that a firm is capable of remaining at the regulatory requirement for all times through liquidations alone, i.e.,  $\theta(t) \geq \theta_{\min}$ for all times $t$, we can drop the indicator function on the firm's capital being positive in $\dot\theta(t)$ as it is always satisfied for $\theta(t) \geq \theta_{\min}$.  Thus by solving for $\dot \theta(t) = 0$ (with the indicator function in \eqref{eq:dot-theta} set equal to $1$), we can conclude that:
\begin{align}
\label{eq:dot-gamma} \dot \Gamma(t) &= -\frac{\dot q(t) [s - \Gamma(t)][\alpha(\bar p - x - \Psi(t) - \ell) + \alpha_{\ell}\ell]}{\alpha q(t) ([s - \Gamma(t)]q(t) - [\bar p - x - \Psi(t) - \ell])} \ind_{\{\theta(t) \leq \theta_{\min}\}}.
\end{align}

For notational simplicity, we will construct the mapping:
\begin{align*}
Z(t,\Gamma(t),q(t),\Psi(t)) &= \frac{[s - \Gamma(t)][\alpha(\bar p - x - \Psi(t) - \ell) + \alpha_{\ell}\ell]}{\alpha q(t) ([s - \Gamma(t)]q(t) - [\bar p - x - \Psi(t) - \ell])} \ind_{\{\theta(t) \leq \theta_{\min}\}}.
\end{align*}
In fact, by the monotonicity of the inverse demand function, we further have that a firm will remain at the $\theta(t) = \theta_{\min}$ boundary for any time $t \geq \tau = \inf\{t \in [0,T] \; | \; \theta(t) \leq \theta_{\min}\}$ provided it does not run out of illiquid assets to sell. Therefore, by solving for the price as a function of liquidations for the equation $\theta(t) = \theta_{\min}$, we find that 
\begin{equation}\label{eq:q-direct}
q(t) = \frac{\bar p - x - \Psi(t) - \ell}{(1 - \alpha \theta_{\min})(s - \Gamma(t))} + \frac{\alpha_{\ell}\theta_{\min}}{1 - \alpha\theta_{\min}}\ell \in \bbr_{++}
\end{equation}
 for $t \geq \tau$.  This provides the price directly as a function of the bank's book.

With the representation of the price $q(t)$ from \eqref{eq:q-direct}, we rearrange terms to find that
\[[s-\Gamma(t)]q(t) - [\bar p - x - \Psi(t) - \ell] = \frac{\alpha\theta_{\min}[\bar p - x - \Psi(t) - \ell] + \alpha_{\ell}\theta_{\min}\ell}{1-\alpha\theta_{\min}}\]
for any time $t \geq \tau$ (equivalently if $\theta(t) \leq \theta_{\min}$).
Therefore we can rewrite $Z$ to only depend on time and the liquidations via
\begin{align*}
Z(t,\Gamma) &= \frac{(1-\alpha\theta_{\min}) [s - \Gamma]}{\alpha\theta_{\min} f_t(t)f_\Gamma(\Gamma)} \ind_{\{t \geq \tau\}}.
\end{align*}
In fact, we can decouple $\dot \Gamma(t)$ from $\dot q(t)$ and thus consider $q(t) = f_t(t)f_\Gamma(\Gamma(t))$ directly and $\dot\Gamma(t)$ to solve the differential equation:
\begin{align}
\label{eq:dot-gamma2}
\dot \Gamma(t) &= -\frac{Z(t,\Gamma(t)) f_t'(t)f_\Gamma(\Gamma(t))}{1 + Z(t,\Gamma(t)) f_t(t) f_\Gamma'(\Gamma(t))}
\end{align}

\begin{remark}\label{rem:purchase}
Of particular interest is that $\dot \Gamma(t) \not\geq 0$ in general.  By Assumption~\ref{ass:idf}, we have that $f_t'(t),f_\Gamma'(\Gamma) \leq 0$ for all times $t$ and liquidations $\Gamma$.  Using the prior computations, as previously discussed for any time $t \geq \tau$, we can conclude that $Z(t,\Gamma(t)) \geq 0$.  Therefore $\dot \Gamma(t) = -\frac{Z(t,\Gamma(t)) f_t'(t)f_\Gamma(\Gamma(t))}{1 + Z(t,\Gamma(t)) f_t(t)f_\Gamma'(\Gamma(t))} \geq 0$ if and only if $f_t(t)f_\Gamma'(\Gamma(t)) \geq -\frac{1}{Z(t,\Gamma(t))}$, otherwise $\dot \Gamma(t) < 0$ and the bank will \emph{purchase} assets at the given price $q(t)$.  As both financial theory and practice indicate such purchasing does not occur in times of a crisis, we utilize the following results in order to calibrate the risk-weights of our model so as to appropriately consider fire sales.  
\end{remark}
Formally, as above, let $\tau := \inf\{t \; | \; \theta(t) \leq \theta_{\min}\} = \inf\{t \; | \; f_t(t) \leq \bar q\}$ be the first time the firm hits the regulatory boundary.

\begin{lemma}\label{lemma:monotonic}
Let the inverse demand function $f_\Gamma$ be such that $(s-\Gamma)f_\Gamma'(\Gamma)/f_\Gamma(\Gamma) \leq 0$ is nondecreasing
for all $\Gamma \in [0,s)$.  If $\alpha \in (-\frac{s f_\Gamma'(0)}{(1-s f_\Gamma'(0))\theta_{\min}} , \frac{1}{\theta_{\min}})$ then
any solution $\Gamma: [\tau,T] \to \bbr$ of~\eqref{eq:dot-gamma2} is such that $\Gamma(t) \in [0,s)$ and $\dot\Gamma(t) \geq 0$ for all times $t$.
\end{lemma}

\begin{remark}
In the prior lemma we require a monotonicity condition on $\frac{(s-\Gamma)f_\Gamma'(\Gamma)}{f_\Gamma(\Gamma)}$.  This term is the ``equivalent'' marginal change in units held to the price change when $\Gamma$ units are liquidated (with the next marginal unit is liquidated externally).  That is, the firm's wealth drops by the same amount under the marginal change in price as if the firm held $\left|\frac{(s-\Gamma)f_\Gamma'(\Gamma)}{f_\Gamma(\Gamma)}\right|$ fewer illiquid assets in their book.  In this sense, this term provides the number of units needed to be sold at the current price in order to counteract the price movement.  Therefore the assumed monotonicity property implies that the firm need not increase the speed it is selling the illiquid assets solely to counteract its own market impacts.
\end{remark}

\begin{theorem}\label{thm:unique}
Consider the setting of Lemma~\ref{lemma:monotonic} with $\alpha \in (-\frac{s f_\Gamma'(0)}{(1-s f_\Gamma'(0))\theta_{\min}} , \frac{1}{\theta_{\min}})$.  There exists a unique solution $(\Gamma,q,\Psi): [0,T] \to [0,s) \times \bbr_{++} \times [0,\bar p - x)$ to the differential system \eqref{eq:dot-gamma2}, \eqref{eq:dot-q}, and \eqref{eq:dot-psi} (and thus for $\theta$ as well for \eqref{eq:dot-theta}).
\end{theorem}

\begin{remark}
Noting that $-\frac{s f_\Gamma'(0)}{1 - s f_\Gamma'(0)} \in [0,1)$ because $f_\Gamma'(0) \leq 0$ by Assumption~\ref{ass:idf}, we are now able to determine the appropriate risk-weight from Lemma~\ref{lemma:monotonic}, i.e., $\alpha \in (-\frac{s f_\Gamma'(0)}{(1-s f_\Gamma'(0))\theta_{\min}} , \frac{1}{\theta_{\min}})$.  If the risk-weight were set too low, i.e., $\alpha \in [0,-\frac{s f_\Gamma'(0)}{(1-s f_\Gamma'(0))\theta_{\min}})$, then the bank would instead purchase assets to remain at the regulatory threshold rather than liquidating as is expected and observed in practice.  The existence and uniqueness results follow for $\alpha < -\frac{s f_\Gamma'(0)}{(1-s f_\Gamma'(0))\theta_{\min}}$ as well, though we will only focus on the risk-weights that match with reality.  In fact, this lower threshold on the risk-weight $\alpha$ can be viewed as a function to map the illiquidity of the asset (measured by $f_\Gamma'(0)$) to an acceptable risk-weight, rather than choosing based on heuristics.
\end{remark}

\begin{remark}\label{rem:default}
The existence and uniqueness results above state that the firm will never liquidate their entire (tradable) portfolio.  This is \emph{not} be the case if the liquid or untradable assets were decreasing in value over time as well; in that scenario, the firm can run out of assets to liquidate.  As the liquidation dynamics (up until the time that the firm becomes completely illiquid), including the existence and uniqueness results, appear similar to the setting stated herein, we focus on the simpler setting in which untradable assets have fixed value over the (short) time horizon $[0,T]$. The value of studying this simpler setting is that it provides ready access to determining the appropriate risk-weight $\alpha$ by capturing the dynamics of a fire sale process before the bank fails without needing to model the failure event as well.
\end{remark}

We will conclude this section by considering two example inverse demand functions $f_\Gamma$: linear and exponential price impacts.  Markets without price impacts is a special case of either inverse demand function by setting $b = 0$.

\begin{example}\label{ex:linear}
Consider the case in which the firm's actions impact the price linearly, i.e., $F(t,\Gamma) = f_t(t)(1 - b\Gamma)$ for $b \in [0,\frac{1}{s})$.  The condition on the inverse demand function for Lemma~\ref{lemma:monotonic} is satisfied for any choice $b \in [0,\frac{1}{s})$.  Further, the risk-weight condition, $\alpha > -\frac{s f_\Gamma'(0)}{(1-s f_\Gamma'(0))\theta_{\min}}$, is satisfied if and only if $\alpha > \frac{sb}{(1+sb)\theta_{\min}}$.  In particular, if $\alpha \geq \frac{1}{2\theta_{\min}}$ then the fire sale situation is always actualized without dependence on the price impact parameter $b$.
\end{example}

\begin{example}\label{ex:exponential}
Consider the case in which the firm's actions impact the price exponentially, i.e., $F(t,\Gamma) = f_t(t)\exp(-b\Gamma)$ for $b \geq 0$.  The condition on the inverse demand function for Lemma~\ref{lemma:monotonic} is satisfied for any choice $b \geq 0$.  Further, the risk-weight condition, $\alpha > -\frac{s f_\Gamma'(0)}{(1-s f_\Gamma'(0))\theta_{\min}}$, is satisfied if and only if $\alpha > \frac{sb}{(1+sb)\theta_{\min}}$.
\end{example}

\subsection{Capital Ratio Requirements in an $n$ Bank System  with Single Representative Tradable Illiquid Asset}\label{sec:n-bank}
Consider the same setting as in Section~\ref{sec:1-bank} but with $n \geq 1$ banks.  Throughout this section we will let firm $i$ have initial banking book defined by $x_i$ units of liquid asset, $s_i$ units of (tradable) illiquid asset, $\ell_i$ units of untradable illiquid asset, and $\bar p_i$ in obligations.  Further, we will consider the (pre-fire sale) market cap for the tradable illiquid asset to be given by $M \geq \sum_{i = 1}^n s_i$.  The inverse demand function will still be assumed to follow Assumption~\ref{ass:idf}.

With the assumption that $\theta_i(0) \geq \theta_{\min}$, we know that firm $i$ will not take any actions unless $\theta_i(t) \leq \theta_{\min}$.  As in the 1 bank case, this first occurs at $\bar q_i = \frac{\bar p_i - x_i - (1-\alpha_{\ell,i}\theta_{\min})\ell_i}{(1 - \alpha \theta_{\min}) s_i}$.  If $\inf_{t \in [0,T]} f_t(t) > \max_i \bar q_i$ then no fire sale occurs.  When a firm does need to take action, we will make the assumption that it is only enough so that the firm remains at the capital ratio requirement.  Thus by solving for $\dot \theta_i(t) = 0$ when $\theta_i(t) \leq \theta_{\min}$ (constructed as in the $n = 1$ bank setting of Section~\ref{sec:1-bank}), we can conclude:
\begin{align*}
\dot \Gamma_i(t) &= -\frac{\dot q(t) [s_i - \Gamma_i(t)][\alpha(\bar p_i - x_i - \Psi_i(t) - \ell_i) + \alpha_{\ell,i}\ell_i]}{\alpha q(t) ([s_i - \Gamma_i(t)]q(t) - [\bar p_i - x_i - \Psi_i(t) - \ell_i])} \ind_{\{\theta_i(t) \leq \theta_{\min}\}}
\end{align*}
with $\dot q(t) = f_t'(t)f_\Gamma(\sum_{i = 1}^n \Gamma_i(t)) + \left[\sum_{i = 1}^n \dot \Gamma_i(t)\right] f_t(t)f_\Gamma'(\sum_{i = 1}^n \Gamma_i(t))$ and
\begin{equation}
\label{eq:dot-psi-n} \dot \Psi_i(t) = \dot \Gamma_i(t) q(t).
\end{equation}

As in the prior section (after consideration of how the prices must evolve so that the firms remain at the required capital ratio), let us consider the mapping
\[Z_i(t,\Gamma) = \frac{(1-\alpha\theta_{\min})[s_i - \Gamma_i(t)]}{\alpha\theta_{\min}f_t(t)f_\Gamma(\sum_{j = 1}^n \Gamma_j)} \ind_{\{\theta_i(t) \leq \theta_{\min}\}}.\]
With this mapping, we can consider the joint differential equation of $\Gamma$ and $q$:
\begin{align}
\label{eq:dot-gamma2-n} \dot \Gamma(t) &= -\left(I + \left(Z(t,\Gamma(t)) \vec{1}^\T\right) f_t(t)f_\Gamma'(\sum_{j = 1}^n \Gamma_j(t))\right)^{-1} \left(Z(t,\Gamma(t)) f_t'(t)f_\Gamma(\sum_{j = 1}^n \Gamma_j(t))\right)\\
\label{eq:dot-q2-n} \dot q(t) &= \frac{f_t'(t)f_\Gamma(\sum_{i = 1}^n \Gamma_i(t))}{1 + \left[\sum_{i = 1}^n Z_i(t,\Gamma(t))\right] f_t(t)f_\Gamma'(\sum_{i = 1}^n \Gamma_i(t))}
\end{align}
where $\vec{1} := (1,1,\dots,1)^\T \in \bbr^n$.

Let $\tau_0 = 0$, $\tau_{k+1} := \inf\{t \in [\tau_k,T] \; | \; \exists i: \; \theta_i(t) \leq \theta_{\min}, \; \theta_i(\tau_k) > \theta_{\min}\}$, and $\tau_{n+1} = T$.
For the remainder, we will order the banks so that $\bar q_i \geq \bar q_{i+1}$ for every $i$.  Due to the monotonicity properties this implies that bank $k$ hits the regulatory threshold only after the first $k-1$ banks.
\begin{lemma}\label{lemma:n-monotonic}
Let the inverse demand function $f_\Gamma$ be such that 
$(M - \Gamma)f_\Gamma'(\Gamma)/f_\Gamma(\Gamma) \leq 0$ is nondecreasing 
for any $\Gamma \in [0,M)$.
If $\alpha \in (-\frac{M f_\Gamma'(0)}{(1-M f_\Gamma'(0))\theta_{\min}} , \frac{1}{\theta_{\min}})$ then 
any solution $\Gamma: [0,T] \to \bbr^n$ of~\eqref{eq:dot-gamma2-n} is such that $\Gamma(t) \in [0,s)$, $\dot\Gamma(t) \in \bbr^n_+$, and $\dot q(t) \leq 0$ for all times $t$.
\end{lemma}

Using this result on monotonicity of the processes, we are able to determine a result on the existence and uniqueness of the system under financial contagion.
\begin{corollary}\label{cor:n-unique}
Consider the setting of Lemma~\ref{lemma:n-monotonic} with $\alpha \in (-\frac{M f_\Gamma'(0)}{(1-M f_\Gamma'(0))\theta_{\min}} , \frac{1}{\theta_{\min}})$.  There exists a unique solution $(\Gamma,q,\Psi): [0,T] \to [0,s) \times \bbr_{++} \times [0,\bar p - x)$ to the differential system \eqref{eq:dot-gamma2-n}, \eqref{eq:dot-q2-n}, and \eqref{eq:dot-psi-n} (and thus for $\theta$ as well). 
\end{corollary}

\begin{remark}\label{rem:nonunique}
As in the single bank $n = 1$ setting, we can consider a situation in which the risk-weight was set too low.  Under such parameters eventually one bank \emph{may} begin purchasing assets rather than liquidating in order to satisfy the capital requirements.  Existence of a solution would still exist in this setting for the $n$ bank case, but uniqueness will no longer hold.
\end{remark}

\begin{remark}
We wish to extend on the comment of Remark~\ref{rem:default} to the setting with $n$ banks.  As in the single firm setting, if the nontradable assets have decreasing value over time due to the financial shock then firms may run out of liquid assets to sell and thus can fail.  
Therefore the strong result that all banks survive for all time presented in, e.g., Corollary~\ref{cor:n-unique} only holds in the special case that the nontradable assets have constant value over time.
As with the single asset setting considered in Remark~\ref{rem:default}, at times between bank failures the system dynamics behave as described in this work; after a bank failure the system parameters would update appropriately (from possible default contagion), then the continuous fire sale model presented herein would begin again until either the terminal time $T$ was reached or another bank failed. 
\end{remark}

\subsection{Capital Ratio Requirements in an $n$ Bank System with $m$ Tradable Illiquid Assets}\label{sec:m-asset}
Consider the same setting as in Section~\ref{sec:1-bank} but with $n \geq 1$ firms and $m \geq 1$ tradable illiquid assets.  Throughout this section we will assume that firm $i$ liquidates its tradable assets in proportion to its holdings $s_i \in \bbr^m_+$.  Notationally, we denote the proportion of assets liquidated by bank $i$ at time $t$ is given by $\Pi_i(t)$.  In this way we can define the vector of total liquidations is given by $\Gamma_i(t) = s_i \Pi_i(t)$.  As in the prior section, we will consider the (pre-fire sale) market cap for the $k^{\text{th}}$ tradable illiquid asset to be given by $M_k \geq \sum_{i = 1}^n s_{ik}$.  The inverse demand function for each asset will still be assumed to follow Assumption~\ref{ass:idf}, i.e., asset $k$ has inverse demand function $F_k(t,\Gamma_k) := f_{t,k}(t)f_{\Gamma,k}(\Gamma_k)$ for any time $t$ and asset liquidations $\Gamma_k$.  We will often consider the vector of inverse demand functions $f_t(t),f_\Gamma(\Gamma) \in \bbr^m_{++}$ to simplify notation.

As in the prior sections, we can construct the derivative of $\theta_i$ over time in order to determine the necessary liquidations so that all firms satisfy the capital ratio requirement.  Using the same logic as above, we can consider the joint differential equation for the fractional liquidations $\Pi$, the vector of prices $q$, and the cash obtained from liquidating tradable assets $\Psi$:
\begin{align}
\label{eq:dot-pi-nm} \begin{split} \dot \Pi(t) &= -\left(I + Z(t,\diag[\Pi(t)]s) \diag[f_t(t)]\diag[f_\Gamma'(s^\T \Pi(t))] s^\T\right)^{-1}\\ &\qquad\qquad \times Z(t,\diag[\Pi(t)]s) \diag[f_t'(t)] f_\Gamma(s^\T \Pi(t))\end{split}\\
\label{eq:dot-q-nm} \dot q(t) &= 
\left(I + \diag[f_t(t)]\diag[f_\Gamma'(s^\T \Pi(t))] s^\T Z(t,\diag[\Pi(t)]s)\right)^{-1} \diag[f_t'(t)]f_\Gamma(s^\T \Pi(t))\\
\label{eq:dot-psi-nm} \dot \Psi(t) &= \diag[\dot \Pi(t)] s q(t)\\
Z(t,\Gamma) &= \diag\left[\ind_{\{\theta(t) \leq \theta_{\min}\}}\right] \diag\left[s \diag[\alpha \theta_{\min}] \diag[f_t(t)] f_\Gamma(\Gamma^\T \vec{1})\right]^{-1} (s - \Gamma) (I - \diag[\alpha \theta_{\min}]).
\end{align}

\begin{lemma}\label{lemma:nm-monotonic}
Let the inverse demand function $f_\Gamma$ be such that 
$(M_k - \Gamma_k)f_{\Gamma,k}'(\Gamma_k)/f_{\Gamma,k}(\Gamma_k) \leq 0$ is nondecreasing for any $\Gamma_k \in [0,M_k)$ for every asset $k$.
If $\alpha_k \in (-\frac{M_k f_{\Gamma,k}'(0)}{(1-M_k f_{\Gamma,k}'(0))\theta_{\min}} , \frac{1}{\theta_{\min}})$ for every asset $k$ then 
any solution $\Pi: [0,T] \to \bbr^n$ of~\eqref{eq:dot-pi-nm} is such that $\Pi(t) \in [0,1)^n$, $\dot\Pi(t) \in \bbr^n_+$, and $\dot q(t) \in -\bbr^m_+$ for all times $t$.
\end{lemma}

Using this result on monotonicity of the processes, we are able to determine a result on the existence and uniqueness of the system under financial contagion.
\begin{corollary}\label{cor:nm-unique}
Consider the setting of Lemma~\ref{lemma:nm-monotonic} with $\alpha_k \in (-\frac{M_k f_{\Gamma,k}'(0)}{(1-M_k f_{\Gamma,k}'(0))\theta_{\min}} , \frac{1}{\theta_{\min}})$ for every asset $k$.  There exists a unique solution $(\Pi,q,\Psi): [0,T] \to [0,1)^n \times \bbr_{++}^m \times [0,\bar p - x)$ to the differential system \eqref{eq:dot-pi-nm}, \eqref{eq:dot-q-nm}, and \eqref{eq:dot-psi-nm} (and thus for $\theta$ as well). 
\end{corollary}

\section{Analytical Stress Test Bounds}\label{sec:stresstest}
As described in the proofs of Lemmas~\ref{lemma:monotonic},~\ref{lemma:n-monotonic}, and~\ref{lemma:nm-monotonic}, we are able to determine upper bounds for the number of assets being sold for each firm in the system.  In the following results we will refine these estimates and use this to determine simple analytical worst-case results for the health of the financial system.  As such, given the initial banking book for each firm, a heuristic for the health of the system can be determined with ease.  Mathematically this is provided by Theorem~\ref{thm:bound}.  Following this result, we will present a quick example to demonstrate the value of these bounds to consider a stochastic stress test.  
Throughout, we will be recalling that, in the single asset setting, firm $i$ hits the regulatory threshold $\theta_{\min}$ when $q(t) = \bar q_i$.

For the remainder of this section we will consider decomposition of the capital ratio as undertaken in the proof of Lemma~\ref{lemma:nm-monotonic}.  With this notion we will define the individual price bounds for liquidations as
\[\bar q_i = \frac{\bar p_i - x_i - (1-\alpha_{\ell,i} \theta_{\min}) \ell_i}{\sum_{k = 1}^m (1 - \alpha_k\theta_{\min}) s_{ik}}\]
for any bank $i$.  Note that these thresholds do not depend on the asset being considered.  Without loss of generality, and as previously discussed, we will assume that firms are ordered so that $\bar q_i$ is a nonincreasing sequence.  
We wish to note that the following analytical bounds, while tight for the single asset $m = 1$ setting (see the numerical examples in Section~\ref{sec:casestudy} below), are typically very weak in the $m \geq 2$ asset setting.  However, the heuristic of considering $\bar q$ as a measure of the risk of each firm is one that requires further study in the $m \geq 2$ setting.

\begin{theorem}\label{thm:bound}
Consider the setting of Corollary~\ref{cor:nm-unique} with $n \geq 1$ banks (ordered by decreasing $\bar q$) and $m \geq 1$ assets.
Define approximate hitting times $\tilde\tau_k$ and bounds on the firm behavior $t \mapsto \tilde\Pi(t)$ for $k = 1,...,n$:
\begin{align*}
\tilde\Pi_i(t) &= \max_{l = 1,...,m}\left\{\frac{\tilde\Gamma_{il}^n(t)}{s_{il}} \; | \; s_{il} > 0\right\}\\
\tilde\Gamma_{il}^k(t) &= \begin{cases} \ind_{\{t < \tilde\tau_{kl}\}} \tilde\Gamma_{il}^{k-1}(t) + \ind_{\{t \geq \tilde\tau_{kl}\}}\left[s_{il} - (s_{il} - \tilde\Gamma_{il}^{k-1}(\tilde\tau_{kl}))\left(\frac{f_{t,l}(t)}{f_{t,l}(\tilde\tau_{kl})}\right)^{\frac{1-\alpha_l\theta_{\min}}{\alpha_l\theta_{\min}\tilde\Lambda_{kl}}}\right] &\text{if } i \leq k \\ 
    0 &\text{else}\end{cases}\\
\tilde\tau_{kl} &= \inf\left\{t \in [\tilde\tau_{k-1,l},T] \; \left| \; f_{t,l}(t)f_{\Gamma,l}\left(\sum_{i = 1}^{k-1} \tilde\Gamma_{il}^{k-1}(t)\right) \leq \bar q_k \right.\right\}\\
\tilde\Lambda_{kl} &= 1 + \frac{1-\alpha_l\theta_{\min}}{\alpha_l\theta_{\min}}\left[\sum_{j = 1}^{k}(s_{jl} - \tilde\Gamma_{jl}^{k-1}(\tilde\tau_{kl}))\right]\frac{f_{\Gamma,l}'\left(\sum_{j = 1}^{k-1} \tilde\Gamma_{jl}^{k-1}(\tilde\tau_{kl})\right)}{f_{\Gamma,l}\left(\sum_{j = 1}^{k-1} \tilde\Gamma_{jl}^{k-1}(\tilde\tau_{kl})\right)}
\end{align*}
where $\tilde\tau_{0l} = 0$, $\tilde\tau_{n+1,l} = T$, and $\tilde\Gamma_{il}^0(t) \equiv 0$.  Then $\Pi_i(t) \leq \tilde\Pi_i^n(t)$ for all times $t \in [0,T]$ and all firms $i = 1,...,n$.
\end{theorem}

With this general analytical construction, we now wish to turn our attention to a specific choice of inverse demand function to provide some additional results.  In particular, as noted in Remark~\ref{rem:exponential}, we will choose the exponential inverse demand function considered in Example~\ref{ex:exponential} to deduce exact analytical formulations.  For the remainder of this section we will make use of the Lambert W function $W: [-\exp(-1),\infty) \to [-1,\infty]$, i.e., the inverse mapping of $x \mapsto x\exp(x)$.

\begin{corollary}\label{cor:bound-exp}
Consider the setting of Theorem~\ref{thm:bound}.  Fix asset $k = 1,...,m$.
Further, consider an exponential inverse demand function $f_{\Gamma,k}(\Gamma_k) := \exp(-b_k\Gamma_k)$ as in Example~\ref{ex:exponential} with $b_k \geq 0$.  
The analytical stress test bounds can be explicitly provided for any $i = 1,...,n$:
\begin{align*}
\tilde\Gamma_{ik}^n(t) &= s_{ik}\left(1 - \prod_{j = i}^n \left(\frac{f_{t,k}(t \wedge \tilde\tau_{j+1,k})}{f_{t,k}(t \wedge \tilde\tau_{jk})}\right)^{\frac{1-\alpha_k\theta_{\min}}{\alpha_k\theta_{\min}\tilde\Lambda_{jk}}}\right)\\
\tilde\tau_{ik} &= \begin{cases} f_{t,k}^{-1}\left(\bar q_{1k}\right) & \text{if } i = 1\\ 
    f_{t,k}^{-1}\left(\left[\frac{\tilde\Lambda_{i-1,k}W\left(\frac{\nu_{i-1,k}}{\tilde\Lambda_{i-1,k}}\exp\left(\frac{1-\alpha_k\theta_{\min}}{\alpha_k\theta_{\min}\tilde\Lambda_{i-1,k}}\left[\log\left(\bar q_{ik}\right) + b_k\sum_{j = 1}^{i-1}s_{jk}\right]\right)\right)}{\nu_{i-1,k}}\right]^{\frac{\alpha_k\theta_{\min}\tilde\Lambda_{i-1,k}}{1-\alpha_k\theta_{\min}}}\right) & \text{if } i \in \{2,...,n\}\end{cases}\\
\tilde\Lambda_{ik} &= 1 - b_k\frac{1-\alpha\theta_{\min}}{\alpha\theta_{\min}}\left[\sum_{j = 1}^i s_{jk} \prod_{h = j}^{i-1}\left(\frac{f_{t,k}(\tilde\tau_{h+1,k})}{f_{t,k}(\tilde\tau_{hk})}\right)^{\frac{1-\alpha_k\theta_{\min}}{\alpha_k\theta_{\min}\tilde\Lambda_{hk}}}\right]\\
\nu_{ik} &= \frac{1-\tilde\Lambda_{ik}}{f_{tk}(\tilde\tau_{ik})^{\frac{1-\alpha_k\theta_{\min}}{{\alpha_k\theta_{\min}\tilde\Lambda_{ik}}}}}
\end{align*}
where $\wedge$ denotes the minimum operator.
\end{corollary}

\begin{remark}\label{rem:boundGamma}
The expanded form $\tilde\Gamma_{ik}^n$ provided in Corollary~\ref{cor:bound-exp} holds for any inverse demand function $f_{\Gamma,k}$ and is not dependent on the choice of the exponential form.  However, the forms of $\tilde\tau_{ik}$ and $\tilde\Lambda_{ik}$ are specific to the exponential inverse demand function considered in Corollary~\ref{cor:bound-exp}.  
\end{remark}

This analytical stress test bound has significant value in considering probability distributions.  All results in this paper, up until now, would require Monte Carlo simulations in order to approximate the distribution of the health of the financial system if there is uncertainty in the parameters.  However, with this analytical bound, we are able to determine analytical worst-case distributions that would be almost surely worse than the actualized results due to the results of Theorem~\ref{thm:bound}.  Thus if the system is deemed healthy enough under this analytical results, it would pass the stress test under the true dynamics as well.

\begin{corollary}\label{cor:probability}
Consider the setting of Corollary~\ref{cor:bound-exp} with exponential price response in time $f_{t,l}(t) := \exp(-a_l t\ind_{\{t < T\}} -a_l T\ind_{\{t \geq T\}})$ to hold for every asset $l$.
Consider a probability space $(\Omega,\mathcal{F},\P)$ and let the parameters $a_l$ be random with known joint distribution.
Fix a time $t \in [0,T]$, the distribution of the price $q(t)$ at time $t$ is bounded by:
\begin{align*}
\P(q(t) \geq q^*) &\geq \P\left(a_l \leq \frac{1}{t}\Phi_{k_ll}^{-1}\left(\log(q_l^*) + b_l\sum_{i = 1}^{k_l} s_{il}\right) \; \forall l = 1,...,m\right)\\
\Phi_{k_ll}^{-1}(x) &= \frac{\alpha_l\theta_{\min}\tilde\Lambda_{k_ll}}{1-\alpha_l\theta_{\min}}W\left(\frac{\nu_{k_ll}}{\tilde\Lambda_{k_ll}}\exp\left(\frac{1-\alpha_l\theta_{\min}}{\alpha_l\theta_{\min}\tilde\Lambda_{k_ll}}x\right)\right) - x.
\end{align*}
where $q_l^* \in [\bar q_{k_l+1},\bar q_{k_l})$ for some $k_l = 0,1,...,n$ (where $\tilde\Lambda_{0l} = 1$, $\nu_{0l} = 0$, $\bar q_0 = 1$, and $\bar q_{n+1} = 0$) for every asset $l$.
\end{corollary}

\begin{remark}\label{rem:probability}
We can generalize the bound for any \emph{random} price response in time $f_t$ from Corollary~\ref{cor:probability} by considering
\begin{align*}
\P(q(t) \geq q^*) &\geq \P\left(f_{t,l}(t) \geq \left[\frac{\tilde\Lambda_{k_ll} W\left(\frac{\nu_{k_ll}}{\tilde\Lambda_{k_ll}}\exp\left(\frac{1-\alpha_l\theta_{\min}}{\alpha_l\theta_{\min}\tilde\Lambda_{k_ll}}\left[\log\left(q^*\right) + b_l\sum_{j = 1}^{k_l} s_{jl}\right]\right)\right)}{\nu_{k_ll}}\right]^{\frac{\alpha\theta_{\min}\tilde\Lambda_{k_ll}}{1-\alpha_l\theta_{\min}}} \; \forall l = 1,...,m\right)
\end{align*}
where $q_l^* \in [\bar q_{k_l+1,l},\bar q_{k_l})$ for some $k_l = 1,...,n$ for every asset $l$.  

This result allows us to consider the case for jointly random price response $f_t$ and price impact parameters $b_l \in [0,\frac{\alpha_l\theta_{\min}}{(1-\alpha_l\theta_{\min})M_l})$ with marginal density $g_b$ through an integral representation. 
The upper bound on the price impact parameters $b$ is so as to guarantee the selected risk-weight satisfies the sufficient conditions considered within this work.
\end{remark}

\section{Case Studies}\label{sec:casestudy}
In this section we will consider four numerical case studies to consider implications of the proposed model.  For simplicity, each of these case studies is undertaken with an exponential inverse demand function.  Further, as the untradable assets do not impact the liquidation dynamics, we will consider examples with $\ell = 0$.  The first three of these numerical case studies is limited to the $m = 1$ asset system with a single, representative, asset as in~\cite{braouezec2016risk,braouezec2017strategic} with $\alpha = 1/(2\theta_{\min})$ throughout.  As such, in each example, we limit the price impact parameters so that $b < 1/M$ as discussed in Remark~\ref{rem:probability}.

The case studies are as follows.  First, we will consider a 20 bank system and determine the effects of the market impacts on the health of the financial system.  Second, we will consider a system with random parameters to study a probabilistic stress test.  
Third, we will consider the effects of changing the regulatory capital ratio threshold.  Finally, we will consider the implications of diversification for a 2 bank, 2 asset system. In these numerical examples we will consider both the numerical solutions to the differential system introduced in Section~\ref{sec:model} and the stress test bounds considered in Section~\ref{sec:stresstest}.  

\begin{example}\label{ex:20bank}
Consider a financial system with $n = 20$ banks, a single tradable illiquid $m = 1$ asset, and a crisis that lasts until the terminal time $T = 1$.  Assume that each bank has liabilities $\bar p_i = 1$ and liquid assets $x_i = \frac{2(i-1)}{475}$ for $i = 1,...,20$.  Additionally, each bank is given $s_i = 2$ units of the illiquid asset; accordingly we set the market capitalization $M = \sum_{i = 1}^{20} s_i = 40$.  We will consider the regulatory environment with threshold $\theta_{\min} = 0.10$ and risk-weight $\alpha = \frac{1}{2\theta_{\min}} = 5$.  Finally, we will take the inverse demand function to have an exponential form, i.e., $F(t,\Gamma) = \exp(-at\ind_{\{t < 1\}} - a\ind_{\{t \geq 1\}} - b\Gamma)$ with $a = -\log(0.95) \approx 0.0513$ and varied market impact parameter $b \in [0,\frac{1}{M})$ which satisfies the conditions of Corollary~\ref{cor:n-unique}.  In this example we will demonstrate the nonlinear response that market impacts $b$ introduce to the health of the firms and clearing prices.

First we wish to consider the impact over time that the market impacts can cause.  To do so we compare the asset prices without market impacts ($b = 0$) to those with high market impacts ($b \approx \frac{1}{M}$).  As depicted in Figure~\ref{fig:20bank-time} we see that the prices with and without price impacts are comparable for (approximately) $t \in [0,0.29]$.  After that time the two systems diverge, drastically so after $t \approx 0.80$.  At that point 18 of the 20 firms (90\%) have hit the regulatory threshold and the feedback effects of their actions are quite evident.  We wish to note the distinction between this steep drop in the prices to the subtle price drop for $t \in [0,0.29]$ when only the first 3 banks have hit their regulatory threshold.  The times at which the firms hit the regulatory threshold at different liquidity situations (i.e.\ no, medium, and high market impacts) are summarized in Table~\ref{table:20bank}.
\begin{table}[ht]
\centering
\begin{tabular}{|c||c|c||c|c||c|c|}
\cline{2-7}
\multicolumn{1}{c}{~} & \multicolumn{2}{|c||}{$\mathbf{b = 0}$} & \multicolumn{2}{c||}{$\mathbf{b = \frac{0.7}{M}}$} & \multicolumn{2}{c|}{$\mathbf{b = \frac{1}{M+10^{-8}}}$} \\
\hline
\textbf{Firm} & \textbf{Numerical} & \textbf{Bounds} & \textbf{Numerical} & \textbf{Bounds} & \textbf{Numerical} & \textbf{Bounds} \\
\hline
\hline
 1 & 0.0000 & 0.0000 & 0.0000 & 0.0000 & 0.0000 & 0.0000 \\
 2 & 0.0823 & 0.0823 & 0.0794 & 0.0794 & 0.0782 & 0.0782 \\
 3 & 0.1649 & 0.1649 & 0.1562 & 0.1562 & 0.1525 & 0.1525 \\
 4 & 0.2478 & 0.2478 & 0.2305 & 0.2305 & 0.2231 & 0.2231 \\
 5 & 0.3311 & 0.3311 & 0.3023 & 0.3023 & 0.2899 & 0.2899 \\
 6 & 0.4148 & 0.4148 & 0.3715 & 0.3715 & 0.3529 & 0.3529 \\
 7 & 0.4989 & 0.4989 & 0.4381 & 0.4381 & 0.4120 & 0.4120 \\
 8 & 0.5832 & 0.5832 & 0.5021 & 0.5021 & 0.4673 & 0.4673 \\
 9 & 0.6680 & 0.6680 & 0.5636 & 0.5635 & 0.5188 & 0.5187 \\
10 & 0.7531 & 0.7531 & 0.6224 & 0.6223 & 0.5663 & 0.5662 \\
11 & 0.8387 & 0.8387 & 0.6786 & 0.6785 & 0.6100 & 0.6099 \\
12 & 0.9245 & 0.9245 & 0.7322 & 0.7321 & 0.6498 & 0.6496 \\
13 &   --   &   --   & 0.7832 & 0.7830 & 0.6856 & 0.6853 \\
14 &   --   &   --   & 0.8315 & 0.8313 & 0.7175 & 0.7172 \\
15 &   --   &   --   & 0.8771 & 0.8770 & 0.7454 & 0.7450 \\
16 &   --   &   --   & 0.9201 & 0.9199 & 0.7694 & 0.7689 \\
17 &   --   &   --   & 0.9605 & 0.9602 & 0.7894 & 0.7888 \\
18 &   --   &   --   & 0.9981 & 0.9978 & 0.8054 & 0.8046 \\
19 &   --   &   --   &   --   &   --   & 0.8173 & 0.8164 \\
20 &   --   &   --   &   --   &   --   & 0.8252 & 0.8242 \\
\hline   
\end{tabular}
\caption{Example~\ref{ex:20bank}: Summary of times at which different firms hit the regulatory threshold $\theta_{\min}$ in the full simulation and in the analytical stress test bounds over no price impacts ($b = 0$), mid-level price impacts ($b = 0.7/M$), and high price impacts ($b \approx 1/M$).}
\label{table:20bank}
\end{table}

With the notion of how high market impacts effect the prices over time, and how the feedback effects can cause virtual jumps in the price, we now wish to consider these effects in more detail by studying only the final state of the system.  In Figure~\ref{fig:20bank-impact} we see that, as more banks hit the threshold capital ratio, the range of price impact thresholds that match that state shrink.  That is, the system becomes more sensitive to the price impact parameter as more banks are at the regulatory threshold.  This is due to the same feedback effects seen in the high price impact scenario of Figure~\ref{fig:20bank-time}.  Further, we see that until about 90\% of the banks (18 out of the 20 firms) hit the regulatory threshold (at about $b \lesssim 0.7/M$), the terminal price is principally affected by the price change in time ($f_t(1) = 0.95$).  At market impacts above this level ($b \gtrsim 0.7/M$) the feedback effects of firm liquidations on each other causes the terminal price to drop drastically.  Thus, providing only a small amount of liquidity to the market can have outsized effects on the health of the system by decreasing the price impacts, though this type of response to a financial crisis would have quickly decreasing marginal returns as evidenced by Figure~\ref{fig:20bank-impact}.

Finally, we wish to consider the analytical stress test bounds.  We see the response of the stress test bounds in the high market impact scenario ($b \approx \frac{1}{M}$) in Figure~\ref{fig:20bank-time}.  This is not depicted in the setting without market impacts as there is no distinction between the exact price process and the bounded price process in this case.  In the high market impact scenario, we see that the exact price process and the stress test bounds provide virtually indistinguishable results for the $t \in [0,0.84]$.  After that time the stress test bound results in a significantly larger shock than the real solution.  As seen in Table~\ref{table:20bank}, the times that firms hit the regulatory threshold are robust between the exact numerical solution and the analytical approximations in all market impact environments (no market impacts, medium impacts, and high impacts).  Finally, we see that the terminal health of the system is replicated with extreme accuracy so long as the price impacts are $b \in [0,\frac{0.9}{M}]$.
\begin{figure}[h!]
\centering
\begin{subfigure}[t]{.47\textwidth}
\includegraphics[width=\textwidth]{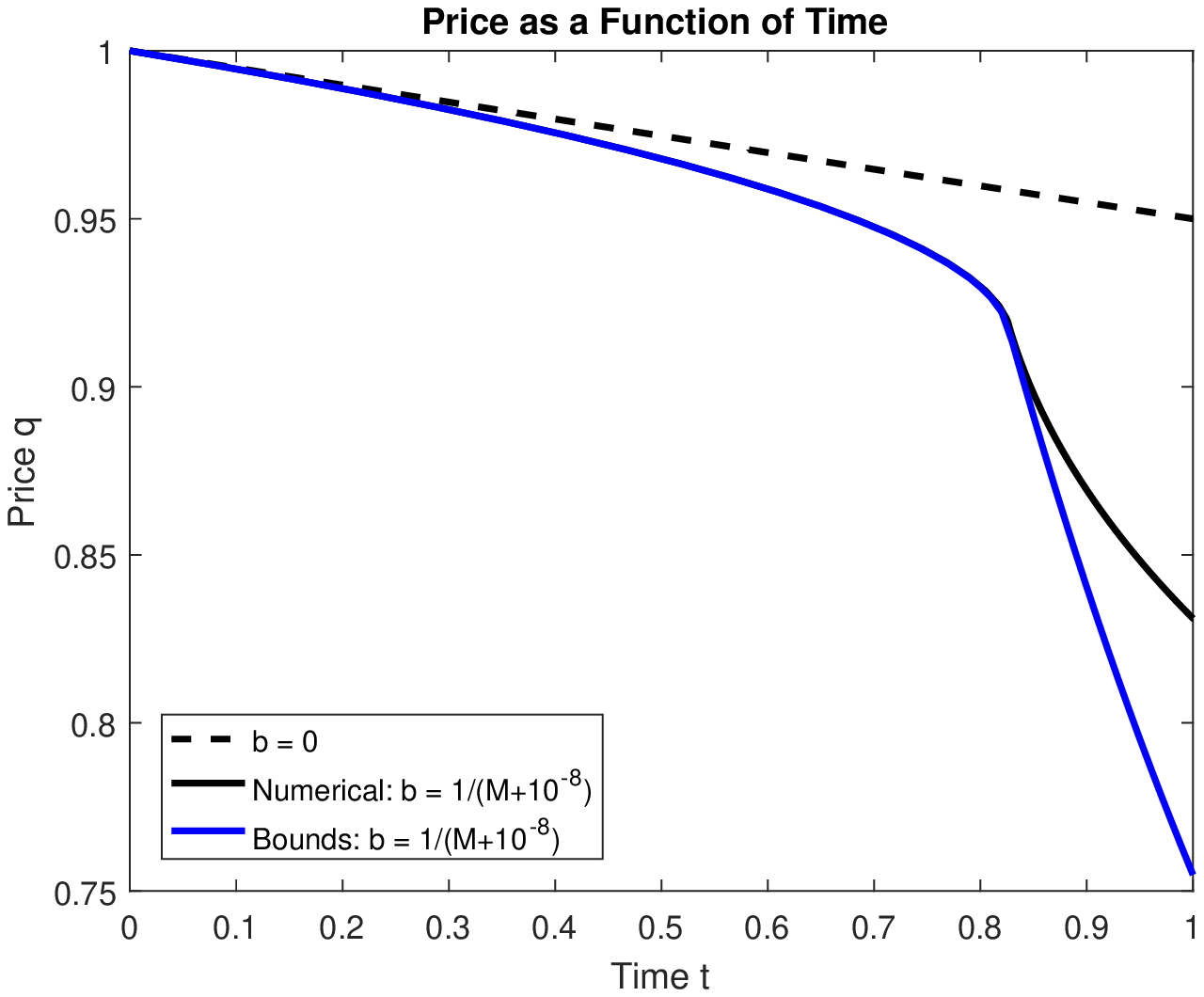}
\caption{A comparison of the asset price over time in a low $b = 0$ and high $b = \frac{1}{M+10^{-8}}$ market impact environment.}
\label{fig:20bank-time}
\end{subfigure}
~
\begin{subfigure}[t]{.47\textwidth}
\includegraphics[width=\textwidth]{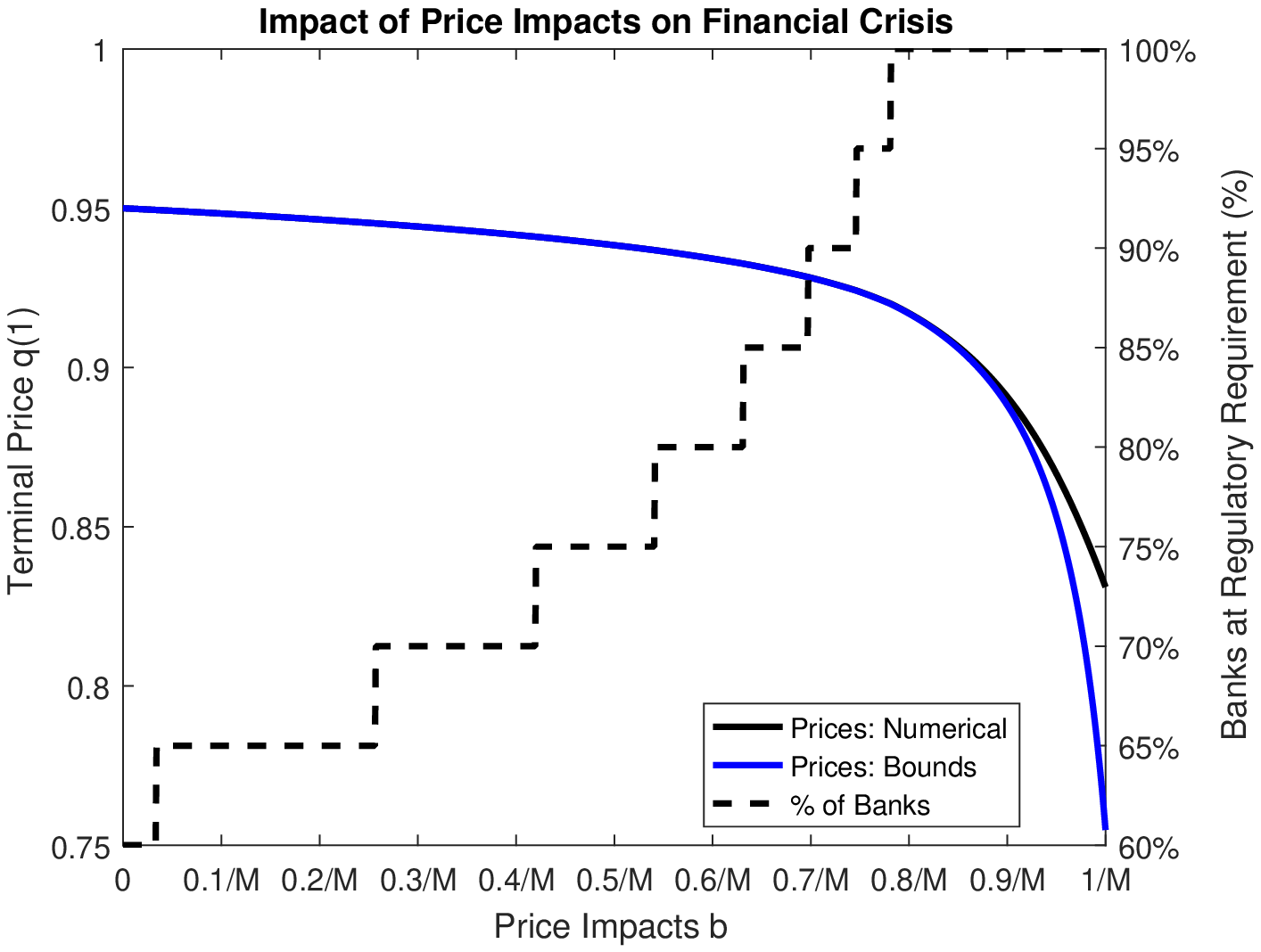}
\caption{The final price $q(1)$ and percentage of firms that have hit their regulatory threshold as a function of the price impact parameter $b$.}
\label{fig:20bank-impact}
\end{subfigure}
\caption{Example~\ref{ex:20bank}: The effects of price impacts on market response in a 20 bank system under the exact differential equation and the analytical stress test bounds.}
\label{fig:20bank}
\end{figure}
\end{example}

\begin{example}\label{ex:20probability}
Consider the setting of Example~\ref{ex:20bank} with exponential inverse demand function $F(t,\Gamma) = \exp(-at\ind_{\{t < 1\}} - a\ind_{\{t \geq 1\}} - b\Gamma)$ for $a \sim \text{Exp}(\mu)$, $\mu = \frac{\log(20)}{\log(20)-\log(19)} \approx 58.4$, and $b = \frac{0.9}{M}$.  The choice of the exponential distribution for $a$ with parameter $\mu$ is so that $\P(F(1,0) \leq 0.95) = 0.05$.  We wish to compare the true distribution for $q(1)$ to the analytical stress test bound given in Corollary~\ref{cor:probability}.  In comparison to the analytical cumulative distribution function given in Corollary~\ref{cor:probability}, the true distribution was found numerically through repeated computation on a (log scaled) regular interval.  Figure~\ref{fig:probability} displays the cumulative distribution functions $\P(q(t) \leq q^*)$ without market impacts (black dashed line), with market impacts (black solid line), and the analytical stress test bound (blue solid line).  Notably, the analytical bound, as seen in Figure~\ref{fig:cdf}, is a very accurate estimate of the true distribution while the market without price impacts distinctly underestimates large price drops.  This is more pronounced in Figure~\ref{fig:cdf-zoom}, which is the same figure but focused on the region for $q^* \in [0.8,0.9]$.  Here we can see that the true distribution is bounded by the analytical stress test distribution, but gives a distribution significantly above the market without price impacts.  In particular, without market impacts the probability $\P(f_t(1) \leq 0.9) \approx 0.002$ whereas $\P(q(1) \leq 0.9) \approx \P(f_t(t)f_\Gamma(\sum_{i = 1}^n \tilde\Gamma_i^n(1)) \leq 0.9) \approx 0.055$.  On the other end, without market impacts the probability $\P(f_t(1) \leq 0.8) \approx 0$ whereas $\P(q(1) \leq 0.8) \approx 0.014$ and $\P(f_t(1)f_\Gamma(\sum_{i = 1}^n \tilde\Gamma_i^n(1)) \leq 0.8) \approx 0.02$.  Thus the analytical stress test is a bound for the true distribution, but an accurate one (as seen in Figure~\ref{fig:cdf}).

\begin{figure}[h!]
\centering
\begin{subfigure}[t]{.47\textwidth}
\includegraphics[width=\textwidth]{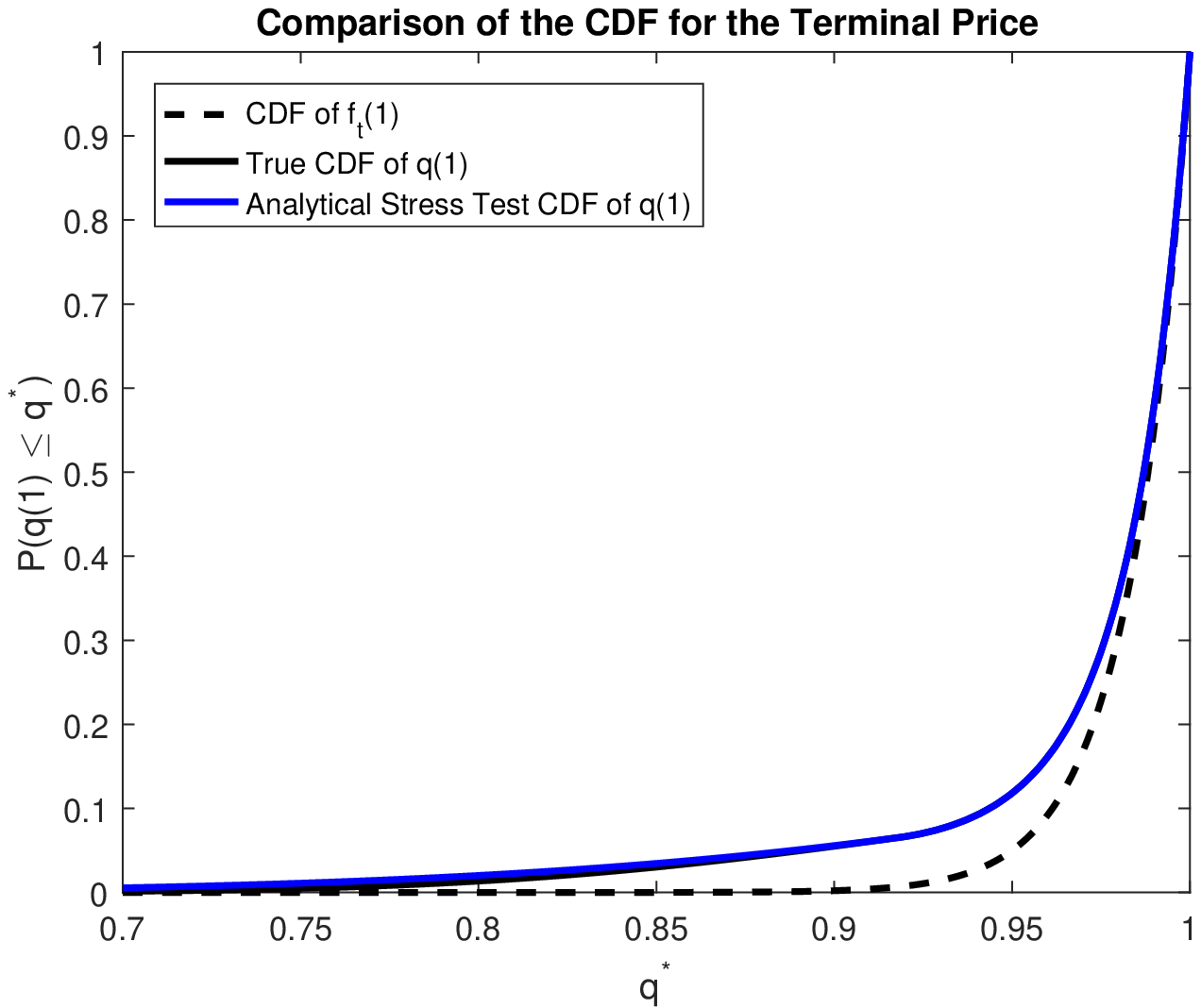}
\caption{The distribution of terminal prices $q(1)$ with and without price impacts.}
\label{fig:cdf}
\end{subfigure}
~
\begin{subfigure}[t]{.47\textwidth}
\includegraphics[width=\textwidth]{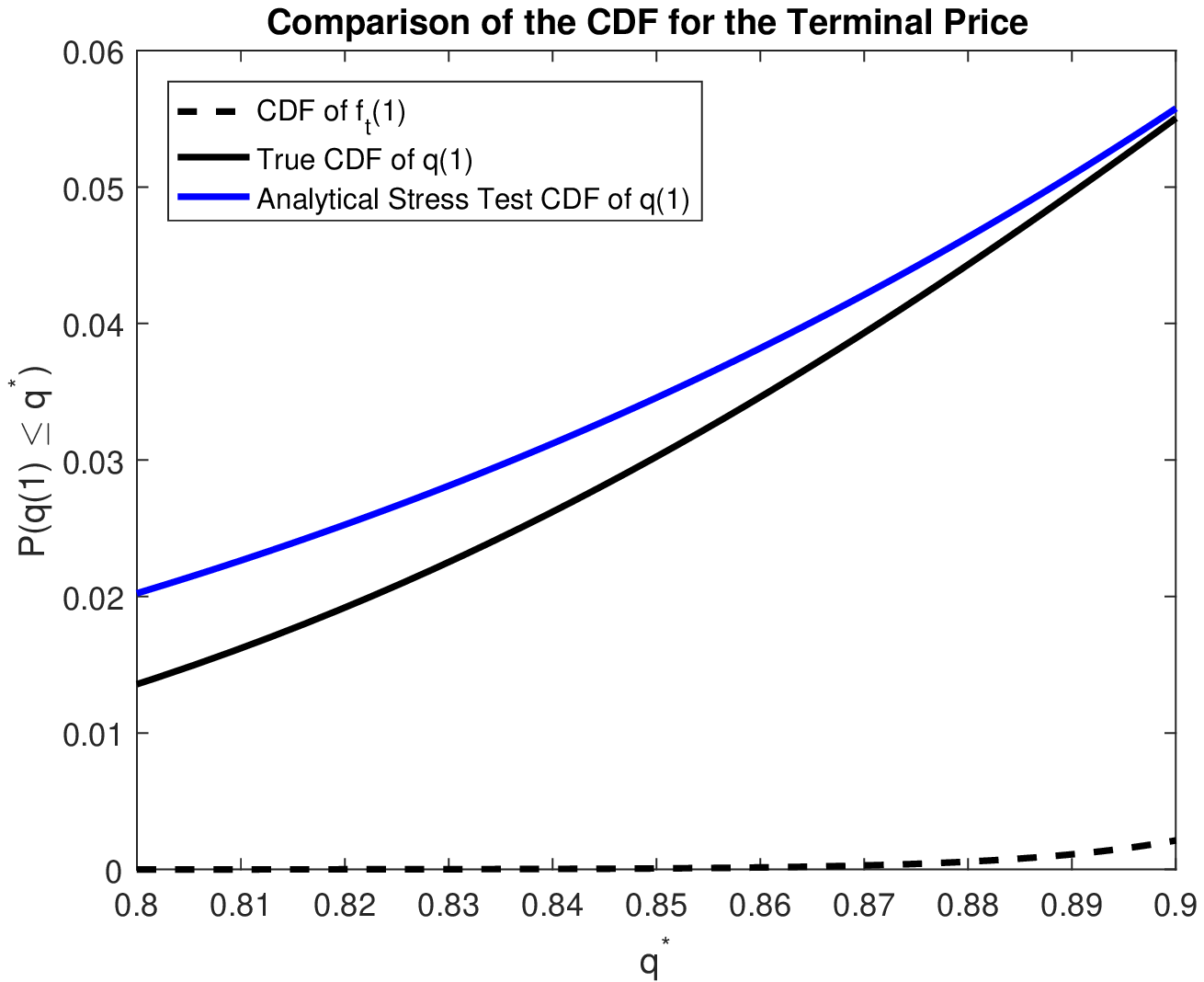}
\caption{Zoomed in view of the distribution of terminal prices $q(1)$.}
\label{fig:cdf-zoom}
\end{subfigure}
\caption{Example~\ref{ex:20probability}: True and analytical stress test distributions for the terminal price $q(1)$ under a randomly stressed financial system of 20 banks.}
\label{fig:probability}
\end{figure}

While considering the probabilistic setting, we can also consider and plot the response to varying the stress scenario given by $a$.  This is depicted in Figure~\ref{fig:stress} by plotting the terminal price $q(1)$ as a function of the price without market impacts $f_t(1)$. The setting without market impacts is the diagonal line by definition.  Market impacts cause feedback effects that drive the price below $f_t(1)$.  All settings coincide for low stress scenarios ($f_t(1) \gtrsim 0.98$) as few banks are driven to the regulatory threshold.  Further, the analytical stress test bound is demonstrably worse than the numerical terminal value for most stresses; however, these occur at typically unrealistic stresses. 
\begin{figure}[h!]
\centering
\includegraphics[width=0.5\textwidth]{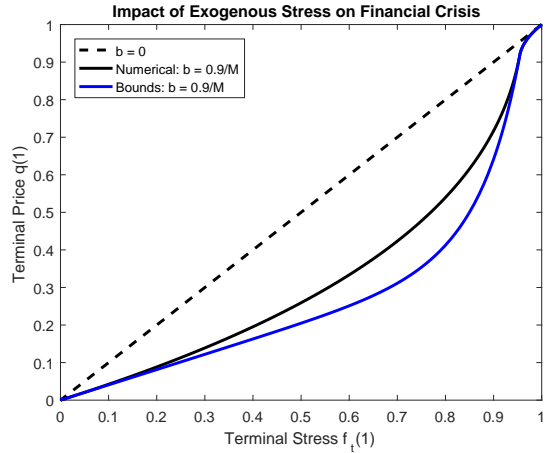}
\caption{Example~\ref{ex:20probability}: The impacts of the stress scenario on market response in a 20 bank system under the exact differential equation and the analytical stress test bounds.}
\label{fig:stress}
\end{figure}
\end{example}

\begin{example}\label{ex:leverage}
Consider a single bank ($n = 1$) and single asset ($m = 1$) system with crisis that lasts until terminal time $T = 1$.  For simplicity, assume that this bank holds no liquid assets, i.e., $x = 0$.  Further, we will directly consider the setting of a leverage constrained firm 
with varying maximal leverage $\lambda_{\max} > 1$.  As we change this leverage requirement, we will assume that the initial banking book for the firm is such that they begin (at time $0$) exactly at the leverage constraint and have a single unit of capital, i.e., $s = \lambda_{\max}$ and $\bar p = \lambda_{\max} - 1$.  For comparison we will fix the inverse demand function to have an exponential form, i.e., $F(t,\Gamma) = \exp(-at\ind_{\{t < 1\}} - a\ind_{\{t \geq 1\}} - b\Gamma)$ with $a = -\log(0.95) \approx 0.0513$ and $b = -\frac{\log(0.9)}{1-1/\log(0.9)} \approx 0.0100$ which satisfies the conditions of Theorem~\ref{thm:unique} for $\lambda_{\max} \in (1,1-\frac{1}{\log(0.9)}) \approx (1,10.50)$.  In this example we will demonstrate the nonlinear response that higher leverage has on the firm behavior and health.

In Figure~\ref{fig:1bank-holdings}, we clearly see that if the leverage requirement is nearly $\lambda_{\max} \approx 1$ then, even though the firm has a banking book that is leverage constrained, very few asset liquidations are necessary and the final portfolio is nearly identical to the original portfolio.  However, as the leverage requirement is relaxed the firm must liquidate a larger percentage of their (larger number of) assets, up to nearly 70\% of all assets.  In fact, once the leverage requirement exceeds $7.15$ the firm has a decreasing number of terminal assets as the leverage requirement increases; this is despite the firm having a greater number of initial assets.  Thus the combination of increasing percentage of assets liquidated and increasing number of initial assets as the leverage requirement $\lambda_{\max}$ increases, the terminal prices decrease as the leverage requirement increases (as depicted in Figure~\ref{fig:1bank-price}).
Finally, we notice that the analytical stress test bounds are accurate for $\lambda_{\max} \lesssim 5.5$.  However, for leverage requirements above that threshold the analytical stress test bounds stop performing well, though clearly are a worst-case bound for the health of the financial system.
\begin{figure}[h!]
\centering
\begin{subfigure}[t]{.47\textwidth}
\includegraphics[width=\textwidth]{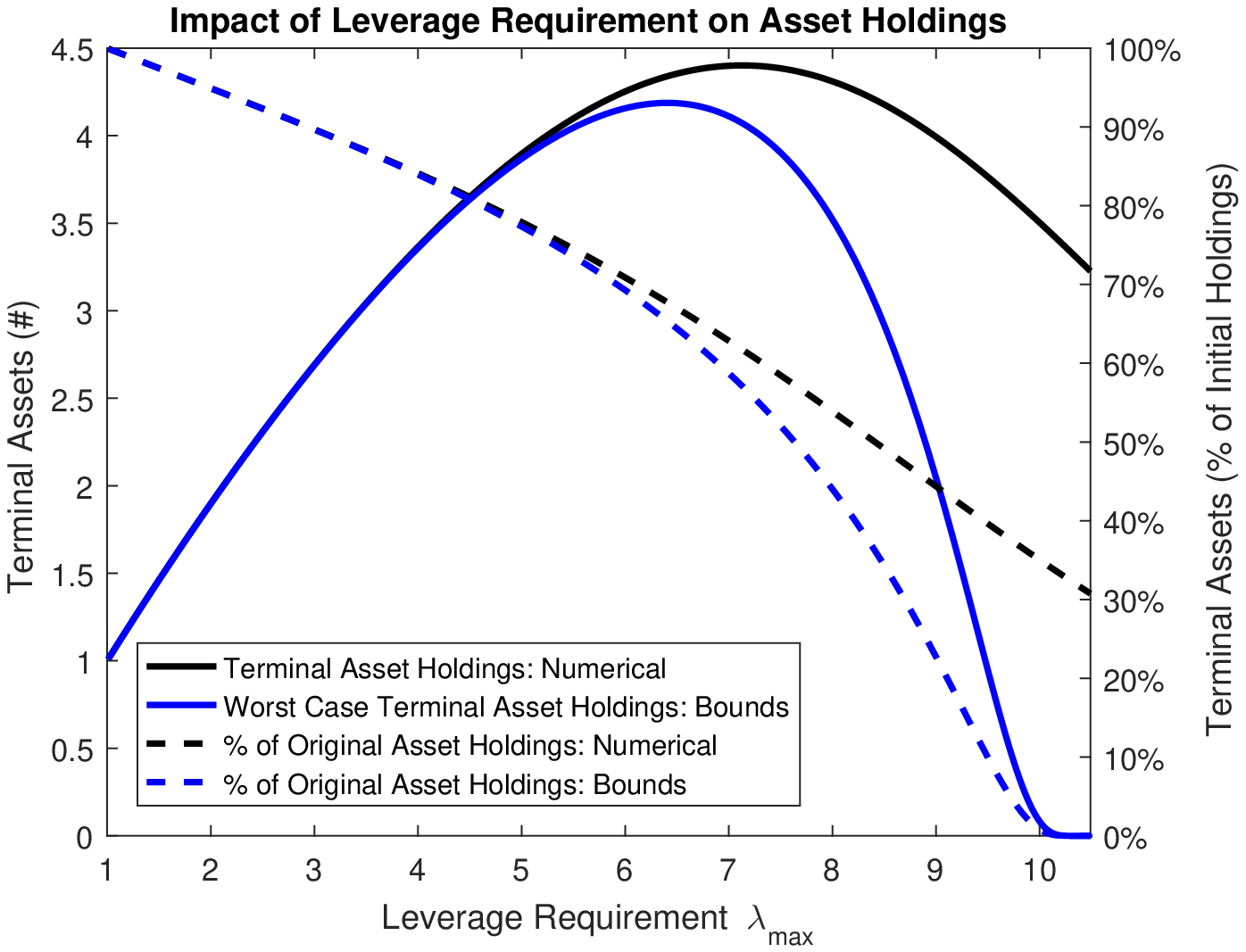}
\caption{The asset holdings and percentage of assets the firm has remaining in its banking book at time $1$ as a function of the leverage constraint $\lambda_{\max}$.}
\label{fig:1bank-holdings}
\end{subfigure}
~
\begin{subfigure}[t]{.47\textwidth}
\includegraphics[width=\textwidth]{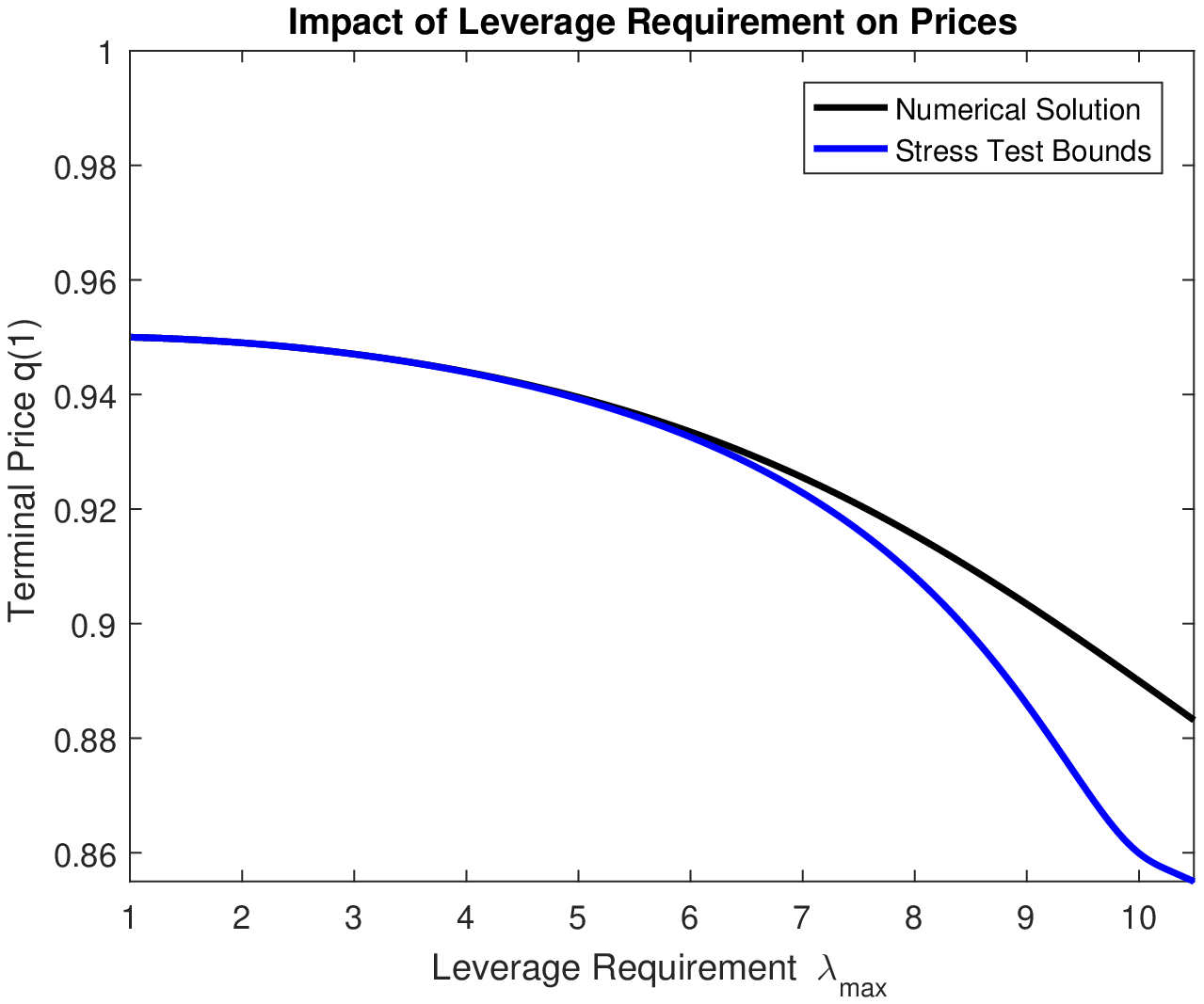}
\caption{The terminal price of the illiquid asset as a function of the leverage constraint $\lambda_{\max}$.}
\label{fig:1bank-price}
\end{subfigure}
\caption{Example~\ref{ex:leverage}: The impacts of the leverage requirement on asset holdings and prices under the exact differential equation and the analytical stress test bounds.}
\label{fig:1bank-leverage}
\end{figure}
\end{example}

\begin{example}\label{ex:2asset}
Consider a two bank ($n = 2$) and two asset ($m = 2$) system with crisis that lasts until terminal time $T = 1$.  For simplicity, assume that both banks hold no liquid assets, i.e., $x = 0$.  Further, assume that both banks have liabilities $\bar p_i = 0.98$ and total initial mark-to-market illiquid assets $s_{i1} + s_{i2} = 2$.  Additionally, we set the market capitalization of each asset to be $M_k = s_{1k} + s_{2k} = 2$.  In this example we will study the implications of diversification by altering the individual portfolios; parameterizing by $\zeta \in [0,2]$, set $s_{11} = (1 - \zeta/2) M_1$, $s_{12} = (\zeta/2) M_2$, $s_{21} = (\zeta/2) M_1$, and $s_{22} = (1 - \zeta/2) M_2$.  We will capture the regulatory environment with threshold $\theta_{\min} = 0.10$ and risk weights $\alpha_1 = \alpha_2 = \frac{1}{2\theta_{\min}} = 5$.  Finally, we will take exponential inverse demand functions
\[F(t,\Gamma) = \left(\exp(-a_1 t \ind_{\{t < 1\}} - a_1 \ind_{\{t \geq 1\}} - b_1 \Gamma_1) \; , \; \exp(-b_2 \Gamma_2)\right)^\T\]
with $a_1 = -\log(0.95) \approx 0.0513$ ($a_2 = 0$) and $b = .4950\times\vec{1}$ (which satisfies the conditions of Corollary~\ref{cor:nm-unique}).  Due to the symmetry of the firms, we will only consider $\zeta \in [0,1]$ for the remainder of this example.  For completeness, the aggregate system (assets $x = 0$, $s = M = (2,2)$, and $\ell = 0$ and liabilities $\bar p = \bar p_1 + \bar p_2 = 1.96$) is considered as well to show the implications of system heterogeneity.

In Figure~\ref{fig:2asset-prices}, we clearly see that diversification of assets does not uniformly improve the market capitalization.  Though the price of asset 1 rises from approximately 0.76 up to nearly 0.87 as the firms become more diversified until perfect diversification ($\zeta = 1$), the price of asset 2 falls from a price of 1 down to approximately 0.92.  The optimal total market capitalization is found at $\zeta = 0.15$, i.e., at a 7.5\%-92.5\% split of assets.  Such a portfolio has very little overlap, thus demonstrating that the contagion effects from holding similar portfolios can easily outweigh the benefits of diversification.  On the other extreme, the completely diverse investment decision ($\zeta = 0$) has the lowest total market capitalization.
Figure~\ref{fig:2asset-liquidation}, demonstrates that the second bank does not start liquidating assets until we reach the optimal $\zeta$ for the market capitalization, i.e., $\Pi_2(1) = 0$ if $\zeta < 0.15$.  In fact, this demonstrates that the contagion effects are exactly those that cause increased diversification to harm the system.
Finally, we note that the aggregated system of this symmetric 2 bank system behaves exactly like the perfectly diversified setting $\zeta = 1$, but can differ greatly in outcome from even a small heterogeneous system.
\begin{figure}[h!]
\centering
\begin{subfigure}[t]{.47\textwidth}
\includegraphics[width=\textwidth]{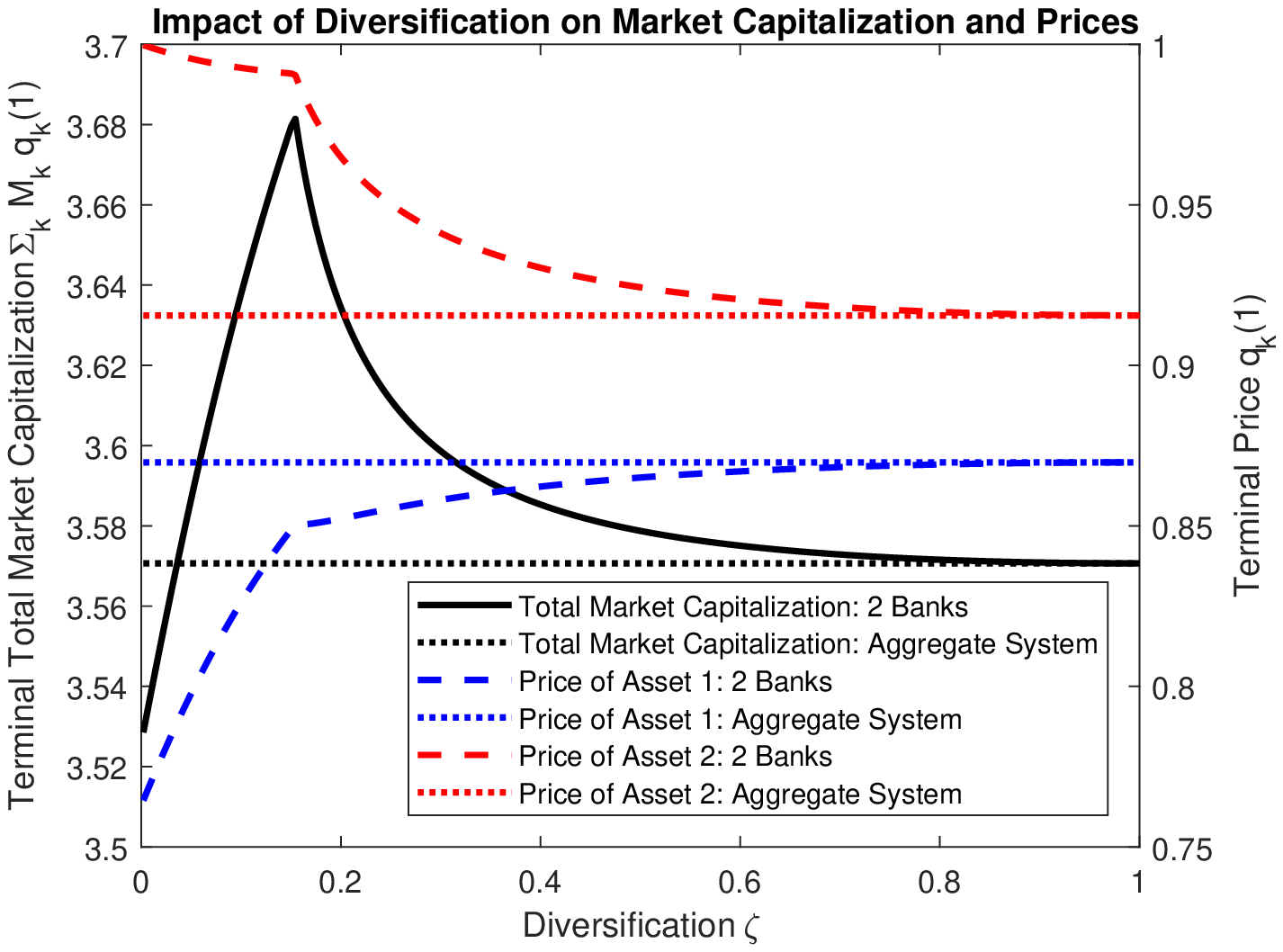}
\caption{The terminal market capitalization and asset prices as a function of portfolio diversity and diversification $\zeta$.}
\label{fig:2asset-prices}
\end{subfigure}
~
\begin{subfigure}[t]{.47\textwidth}
\includegraphics[width=\textwidth]{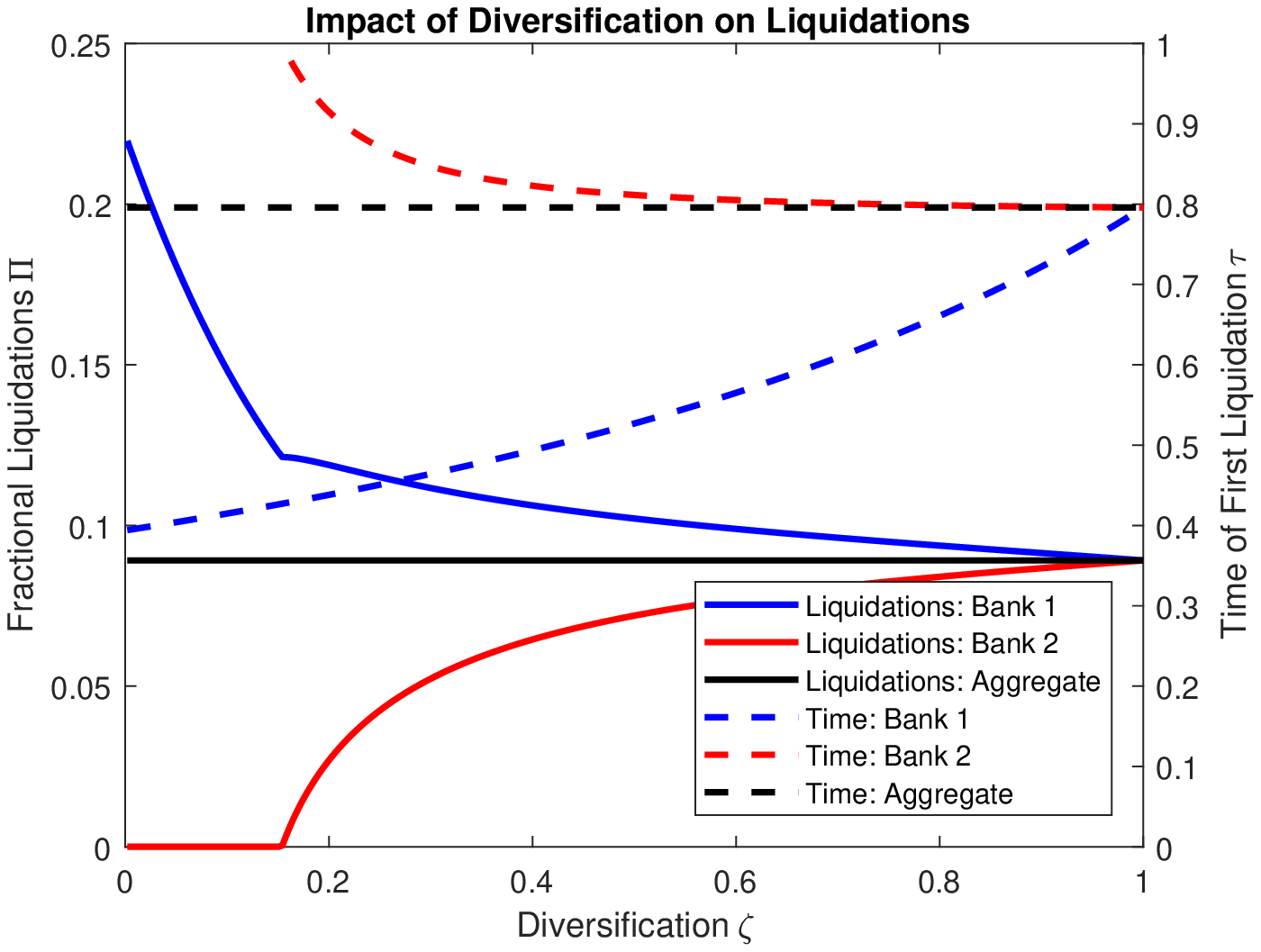}
\caption{The terminal fractional liquidations $\Pi_i(1)$ and first liquidation time $\tau_i$ as a function of portfolio diversity and diversification $\zeta$.}
\label{fig:2asset-liquidation}
\end{subfigure}
\caption{Example~\ref{ex:2asset}: The impacts of portfolio diversity and diversification on asset prices and holdings.}
\label{fig:2asset-diversification}
\end{figure}
\end{example}

\section{Conclusion}\label{sec:conclusion}
In this paper we considered a dynamic model of price-mediated contagion that extends the work of \cite{CFS05,feinstein2016leverage,braouezec2016risk,braouezec2017strategic}.  The focus of this work was on capital ratio requirements and risk-weighted assets.  In analyzing this model, we determine bounds for appropriate risk-weights for an asset that is dependent on the liquidity of the asset itself, as modeled through the price impacts of liquidating the asset.  Under the appropriate risk-weights, we find existence and uniqueness for the firm behavior and system health.  However, though the output of the model can be computed with standard methods, an analytical solution cannot be found; an analytical bound on the health of the system in a stressed scenario was provided.  This analytical stress test bound can be used to analyze random stresses and find the probability for the system health.
We wish to note that an important extension of this model is to more fully consider the setting in which the nontradable assets have prices that drop over time, thus allowing for bank failures.  By determining bank failures over time, properly modeling the failure event and default contagion, and determining the updated system parameters after default (to be run on the current proposed model) would be an interesting exercise to model the reality of a contagious event more closely.

\appendix
\section{Proofs from Section~\ref{sec:1-bank}}
Define the mapping $\Lambda(t) := 1 + Z(t,\Gamma(t)) f_t(t)f_\Gamma'(\Gamma(t))$, which will be utilized throughout many of the following proofs.

\subsection{Proof of Lemma~\ref{lemma:monotonic}}
\begin{proof}
We will demonstrate that if a solution exists then it must satisfy the monotonicity property.  To do so, first, we note that the condition on the risk-weight $\alpha$ is equivalent to $\alpha\theta_{\min} < 1$ and $\Lambda(\tau) > 0$.  Therefore we find that $\dot \Gamma(\tau) > 0$.  Now we wish to show that $\dot \Lambda(t)$ has the same sign as $\dot \Gamma(t)$, i.e., $\dot \Gamma(t)\dot \Lambda(t) \geq 0$.
\begin{align*}
\Lambda(t) &= 1 + Z(t,\Gamma(t)) f_t(t)f_\Gamma'(\Gamma(t))
 = 1 + \frac{1-\alpha \theta_{\min}}{\alpha \theta_{\min}} \frac{s-\Gamma(t)}{f_\Gamma(\Gamma(t))} f_\Gamma'(\Gamma(t))\\
\dot \Lambda(t) &= \frac{1-\alpha\theta_{\min}}{\alpha\theta_{\min}} \frac{d}{dt}\frac{(s-\Gamma(t)) f_\Gamma'(\Gamma(t))}{f_\Gamma(\Gamma(t))}
 = \frac{(1-\alpha\theta_{\min})\dot\Gamma(t)}{\alpha\theta_{\min}} \frac{d}{d\Gamma} \left[\frac{(s-\Gamma)f_\Gamma'(\Gamma)}{f_\Gamma(\Gamma)}\right]_{\Gamma = \Gamma(t)}
\end{align*}
Therefore, by $\alpha\theta_{\min} \in (0,1)$, we find that $\dot\Gamma(t) \dot \Lambda(t)\geq 0$ if and only if
\[\frac{d}{d\Gamma} \left[\frac{(s-\Gamma)f_\Gamma'(\Gamma)}{f_\Gamma(\Gamma)}\right]_{\Gamma = \Gamma(t)} \geq 0\]
which is true by assumption.  We will now use an induction argument to prove $\Lambda(t) > 0$ for all times $t \in [\tau,T]$:
\begin{itemize}
\item At time $\tau$ we have (by assumption) that $\Lambda(\tau) > 0$.
\item For any time $t \in [\tau,T)$ such that $\Lambda(t) > 0$ then it must be that $\Lambda(u) > 0$ for every $u \in [t,t+\epsilon]$ for some $\epsilon > 0$ by continuity of $\Gamma$ and therefore of $\Lambda$ (and that $\Lambda$ is strictly above 0).  
\item For any time $t \in (\tau,T]$ such that $\Lambda(u) > 0$ for every $u \in [\tau,t)$ then $\dot\Gamma(u) > 0$ and, as a consequence, $\dot \Lambda(u) \geq 0$ for every $u \in [\tau,t)$.  This implies $\Lambda(t) > 0$ as well.
\end{itemize}
Therefore, if $\Lambda(\tau) > 0$ it must hold that $\Lambda(t) \geq \Lambda(\tau)$ for all times $t \geq \tau$ (which implies $\dot \Gamma(t) \geq 0$ for all times $t \geq \tau$).  

Finally, we will demonstrate that, if a solution $\Gamma: [0,T] \to \bbr$ exists, then $\Gamma(t) < s$ for all times $t \in [0,T]$.  By definition $\Gamma(t) = 0$ for all times $t \in [0,\tau]$, so we begin with $\Gamma(\tau) = 0$.  Take $T^* = \inf\{t \in [\tau,T] \; | \; \Gamma(t) \geq s\}$ and assume this infimum is taken over a nonempty set.
On $u \in [\tau,T^{*})$ we have that:
\begin{align*}
\dot\Gamma(u) &= -\frac{Z(u,\Gamma(u))f_t'(u)f_\Gamma(\Gamma(u))}{\Lambda(u)} \leq -\frac{(1-\alpha\theta_{\min}) \inf_{t \in [\tau,T^{*}]} f_t'(t)}{\alpha\theta_{\min} f_t(T^{*}) \Lambda(\tau)} (s - \Gamma(u))
\end{align*}
and $\inf_{t \in [\tau,T^*]} f_t'(t)$ is attained as we are infimizing a continuous function over a compact space.  This differential equation implies $\Gamma(u) \leq s\left[1 - \exp\left(\frac{(1-\alpha\theta_{\min}) \inf_{t \in [\tau,T^{*}]} f_t'(t)}{\alpha\theta_{\min} f_t(T^{*}) \Lambda(\tau)}(u-\tau)\right)\right] < s$ for any time $u \in [\tau,T^{*})$.  In particular, by continuity, this implies that $\Gamma(T^{*}) < s$.
\end{proof}

\subsection{Proof of Theorem~\ref{thm:unique}}
\begin{proof}
We will use Lemma~\ref{lemma:monotonic} to prove the existence and uniqueness of a solution $(\Gamma,q,\Psi)$.  First, for all times $t \in [0,\tau]$ there exists a unique solution given by $\Gamma(t) = 0$, $q(t) = f_t(t)$, and $\Psi(t) = 0$.  

Now consider the initial value problem with initial condition at $t = \tau$.  We will consider the differential equation for $\Gamma$ given in \eqref{eq:dot-gamma2}.  As this equation is no longer dependent on either $q$ or $\Psi$ we can consider the existence and uniqueness of the liquidations $\Gamma$ separately.  Indeed, from $\Gamma$, we can define $q(t) = f_t(t)f_\Gamma(\Gamma(t))$ for all times $t$, thus the existence and uniqueness of $\Gamma$ provides the same results for $q$.  The results for $\Psi$ follow from the same logic as $\Gamma$ and thus will be omitted herein. In our consideration of \eqref{eq:dot-gamma2} we will consider a modification of the function $\Lambda(t)$ to be given by $\bar\Lambda(\Gamma)$ so that its dependence on the liquidations is made explicit:
\begin{align*}
\dot \Gamma(t) &= -\frac{(1-\alpha\theta_{\min})(s-\Gamma(t))f_t'(t)}{\alpha \theta_{\min} f_t(t)\bar\Lambda(\Gamma(t))} =: g(t,\Gamma(t)) \quad \text{and} \quad \bar\Lambda(\Gamma) = 1 + \frac{(1-\alpha \theta_{\min})(s-\Gamma)}{\alpha \theta_{\min} f_\Gamma(\Gamma)} f_\Gamma'(\Gamma).
\end{align*}
Now we wish to consider the initial value problem for $\Gamma$ with dynamics given by $g$ and initial value $\Gamma(\tau) = 0$.  
Before continuing we wish to note that the function $\bar\Lambda$ is constant in time, i.e., only depends on the total number of units liquidated $\Gamma$ and not on the time.

Define the domain $U = \left\{\Gamma \in [0,s) \; | \; \bar\Lambda(\Gamma) > \frac{1}{2} \Lambda(\tau) = \frac{1}{2}\left[1 + \frac{(1-\alpha\theta_{\min})s}{\alpha\theta_{\min}} f_\Gamma'(0)\right]\right\}$.
We wish to note from the previous proof that $\Lambda$ is nondecreasing in time, thus any solution must lie in $U$, i.e., it must satisfy $\Gamma(t) \in U$ for all times $t \in [\tau,T]$.  From the definition of $U$ as well as the property that $\bar\Lambda$ is constant in time, we can conclude 
$\alpha\theta_{\min} f_t(t) \bar\Lambda(\Gamma) > \frac{1}{2} \alpha\theta_{\min} f_t(t)\Lambda(\tau) > 0$
for any $\Gamma \in U$ and any time $t \in [\tau,T]$, and thus the denominator in $g$ is always strictly greater than $0$.  From this we can conclude that $g$ and $\dG g$ are continuous mappings over $[\tau,T] \times U$ and thus $\Gamma \in U \mapsto g(t,\Gamma)$ is locally Lipschitz for any time $t \in [\tau,T]$.  This implies there exists some $\delta > 0$ such that $\Gamma: [\tau,\tau+\delta] \to U$ is the unique solution satisfying $\dot \Gamma(t) = g(t,\Gamma(t))$ for all times $t \in [\tau,\tau+\delta]$.  From a sequential application of this approach (i.e., consider now an initial value problem starting at time $\tau+\delta$) we can either conclude that there exists a unique solution over the entire time range $\Gamma: [\tau,T] \to U$ or there exists some maximal domain $[\tau,T^*) \subsetneq [\tau,T]$ over which we can conclude the existence and uniqueness.  We will finish by focusing on this second case to prove a contradiction.  To do this we will first show that $g$ is bounded on $[\tau,T] \times U$.  By definition, we have that $g(t,\Gamma) \geq 0$ for any $t \in [\tau,T]$ and $\Gamma \in U$.  
In fact, we find that
$0 \leq g(t,\Gamma) \leq -\frac{2(1-\alpha\theta_{\min})s \inf_{u \in [\tau,T]} f_t'(u)}{\alpha\theta_{\min} f_t(T) \Lambda(\tau)}$
where $\inf_{u \in [\tau,T]} f_t'(u)$ is attained as this is optimizing a continuous function over a compact space.  With the boundedness of $g$ we find that the limit $\Gamma(T^*) := \lim_{t \nearrow T^*} \Gamma(t)$ exists.  Furthermore, $\bar\Lambda(\Gamma(T^*)) \geq \Lambda(\tau) > \frac{1}{2}\Lambda(\tau)$ and $\Gamma(T^*) < s$  (by the result of Lemma~\ref{lemma:monotonic}).  Thus we can continue our solution to $\Gamma: [\tau,T^*] \to U$ and find a contradiction to $[\tau,T^*)$ being the maximal domain.
\end{proof}

\section{Proofs from Section~\ref{sec:n-bank}}
Throughout this section, without loss of generality, assume that banks are ordered with decreasing $\bar q$.

\subsection{Proof of Lemma~\ref{lemma:n-monotonic}}
\begin{proof}
We will consider this argument by induction.  
In the $n$ bank case, define
\begin{align*}
\Lambda(t,\Gamma) &:= 1 + \sum_{j = 1}^n Z_j(t,\Gamma) f_t(t)f_\Gamma'(\sum_{j = 1}^n \Gamma_j) 
= \operatorname{det}\left[I + \left(Z(t,\Gamma) \vec{1}^\T\right) f_t(t)f_\Gamma'(\sum_{j = 1}^n \Gamma_j)\right].
\end{align*}  
By the ordering of banks and the assumption that no firm will modify its portfolio until it hits the regulatory threshold we know that $\Gamma_i(t) = 0$ if $t < \tau_i$. 
We will consider this proof by induction.  Note first that $\tau_0 = 0$ by construction.  By $\Gamma_i(t) = 0$ if $t < \tau_i$, the results are trivial for $t \in [0,\tau_1)$.  Now take $k \in \{1,2,...,n\}$.
For any time $t \in [\tau_k,\tau_{k+1})$ (or $t \in [\tau_k,T]$ if $\tau_{k+1} \geq T$)
\begin{align*}
\Lambda(t,\Gamma(t)) 
&= 1 + \frac{1-\alpha\theta_{\min}}{\alpha\theta_{\min}}\frac{\left[\sum_{j = 1}^k \left(s_j-\Gamma_j(t)\right)\right]f_\Gamma'(\sum_{j = 1}^k \Gamma_j(t))}{f_\Gamma(\sum_{j = 1}^k \Gamma_j(t))}\\
\dot\Lambda(t,\Gamma(t)) 
&= \frac{1-\alpha\theta_{\min}}{\alpha\theta_{\min}}\frac{d}{d\Gamma}\left[\frac{\left[\left(\sum_{j = 1}^k s_j\right)-\Gamma\right]f_\Gamma'(\Gamma)}{f_\Gamma(\Gamma)}\right]_{\Gamma = \sum_{i = 1}^k \Gamma_i(t)} \sum_{i = 1}^k \dot\Gamma_i(t)
\end{align*}
Using the same logic as in Lemma~\ref{lemma:monotonic}, we recover that $\dot\Lambda(t,\Gamma(t)) \geq 0$ if and only if $\sum_{i = 1}^k \dot\Gamma_i(t) \geq 0$ so long as 
$\frac{d}{d\Gamma}\left[\left(\left[\left(\sum_{j = 1}^k s_j\right)-\Gamma\right]f_\Gamma'(\Gamma)\right)/f_\Gamma(\Gamma)\right] \geq 0$ at $\Gamma = \sum_{i = 1}^k \Gamma_i(t)$.
To prove this sufficient condition, consider the assumptions on the inverse demand function $f_\Gamma$ and assume $\Gamma = \sum_{i = 1}^k \Gamma_i(t) \in [0,\sum_{i = 1}^k s_i) \subseteq [0,M)$:
If $\Gamma$ is such that $f_\Gamma'(\Gamma)^2 \geq f_\Gamma(\Gamma) f_\Gamma''(\Gamma)$ then
\begin{align*}
\frac{d}{d\Gamma}&\left[\frac{\left[\left(\sum_{j = 1}^k s_j\right)-\Gamma\right]f_\Gamma'(\Gamma)}{f_\Gamma(\Gamma)}\right]_{\Gamma = \sum_{i = 1}^k \Gamma_i(t)} 
= \frac{d}{d\Gamma}\left[\frac{\left[M-\Gamma\right]f_\Gamma'(\Gamma)}{f_\Gamma(\Gamma)} - \frac{\left[M - \left(\sum_{j = 1}^k s_j\right)\right]f_\Gamma'(\Gamma)}{f_\Gamma(\Gamma)}\right]_{\Gamma = \sum_{i = 1}^k \Gamma_i(t)}\\
&= \left[\frac{d}{d\Gamma}\frac{\left[M-\Gamma\right]f_\Gamma'(\Gamma)}{f_\Gamma(\Gamma)} - \left[M - \left(\sum_{j = 1}^k s_j\right)\right]\frac{f_\Gamma(\Gamma)f_\Gamma''(\Gamma) - f_\Gamma'(\Gamma)^2}{f_\Gamma(\Gamma)^2}\right]_{\Gamma = \sum_{i = 1}^k \Gamma_i(t)} \geq 0.
\end{align*}
Otherwise $f_\Gamma'(\Gamma)^2 < f_\Gamma(\Gamma) f_\Gamma''(\Gamma)$ and the result follows directly from the construction of the derivative.

Further, by construction, if $\Lambda(\tau_k,\Gamma(\tau_k)) > 0$ then $\dot q(\tau_k) \leq 0$ and $\dot \Gamma(\tau_k)$ exists.
We now want to demonstrate that $\Lambda(\tau_k,\Gamma(\tau_k)) > 0$.  By construction this is true if and only if $\alpha > -\frac{\left(\sum_{i = 1}^{k-1} [s_i - \Gamma_i(\tau_k)] + s_k\right) f_\Gamma'(\sum_{i = 1}^{k-1} \Gamma_i(\tau_k))/f_\Gamma(\sum_{i = 1}^{k-1} \Gamma_i(\tau_k))}{\left(1 - \left(\sum_{i = 1}^{k-1} [s_i - \Gamma_i(\tau_k)] + s_k\right) f_\Gamma'(\sum_{i = 1}^{k-1} \Gamma_i(\tau_k))/f_\Gamma(\sum_{i = 1}^{k-1} \Gamma_i(\tau_k))\right) \theta_{\min}}$.  With $\Gamma_k(\tau_k) = 0$ by definition and by the assumption on the inverse demand function 
$0 \geq \frac{\left(\sum_{i = 1}^k s_i - \sum_{i = 1}^k \Gamma_i(\tau_k)\right) f_\Gamma'(\sum_{i = 1}^k \Gamma_i(\tau_k))}{f_\Gamma(\sum_{i = 1}^k \Gamma_i(\tau_k))} \geq \frac{\left(\sum_{i = 1}^k s_i\right) f_\Gamma'(0)}{f_\Gamma(0)} \geq \frac{M f_\Gamma'(0)}{f_\Gamma(0)} = M f_\Gamma'(0)$.
Therefore, if $\alpha > -\frac{M f_\Gamma'(0)}{(1 - M f_\Gamma'(0))\theta_{\min}}$ then $\Lambda(\tau_k,\Gamma(\tau_k)) > 0$.

Next we wish to show that $\dot\Gamma(t) \in \bbr^n_+$ for all times $t$.  As originally constructed we have that for any $i \leq k$:
$\dot\Gamma_i(t) = \frac{\dot q(t)[s_i - \Gamma_i(t)][\bar p_i - x_i - \Psi_i(t)]}{q(t)[(s_i - \Gamma_i(t))q(t) - (\bar p_i - x_i - \Psi_i(t))]} \ind_{\{\theta_i(t) \leq \theta_{\min}\}}$.
If a bank is brought above the regulatory threshold they will not perform any transactions, i.e., $\dot\Gamma_i(t) = 0$, but this can only occur if $\dot q(t) > 0$.  Otherwise $(1-\alpha\theta_{\min})(s_i - \Gamma_i(t))q(t) = \bar p_i - x_i - \Psi_i(t)$ as the firm will need to remain at the regulatory threshold.  As such we can simplify $\dot\Gamma_i(t)$ as
$\dot\Gamma_i(t) = -\frac{\dot q(t)(1-\alpha\theta_{\min})[s_i - \Gamma_i(t)]}{\alpha\theta_{\min}q(t)}\ind_{\{\theta_i(t) \leq \theta_{\min}\}}$.
This allows us to conclude that $\dot\Gamma_i(t)$ has the opposite sign of $\dot q(t)$, i.e., $\dot\Gamma(t) \in \bbr^n_+$.

Finally, we will now demonstrate that $\Gamma(t) \in [0,s)$ for all times $t \in [\tau_k,\tau_{k+1})$ (or $t \in [\tau_k,T]$ if $\tau_{k+1} \geq T$) by induction for any $k \in \{0,1,...,n\}$.  
As noted above, we find that $\dot q(t) \leq 0$ for all times $t \in [\tau_k,\tau_{k+1})$.  
By assumption $\Gamma(t) \in [0,s)$ for all times $t \in [0,\tau_k]$, so we begin with $\Gamma(\tau_k) \in [0,s)$.  Take $T^* = \inf\{t \in [\tau_k,\tau_{k+1}) \; | \; \exists i: \; \Gamma_i(t) \geq s_i\}$ and assume this infimum is taken over a nonempty set.  On $u \in [\tau_k,T^*)$ we have that:
\begin{align*}
\dot q(u) &= \frac{f_t'(u)f_\Gamma(\sum_{j = 1}^n \Gamma_j(u))}{\Lambda(u)} \geq \frac{\inf_{t \in [\tau_k,T^{*}]} f_t'(t) f_\Gamma(\sum_{j = 1}^n \Gamma_j(u))}{\Lambda(\tau_k,\Gamma(\tau_k))}\\
\dot \Gamma_i(u) &= -\frac{(1-\alpha\theta_{\min})\dot q(u) [s_i - \Gamma_i(u)]}{\alpha\theta_{\min}f_t(u)f_\Gamma(\sum_{j = 1}^n \Gamma_j(u))} \leq -\frac{(1-\alpha\theta_{\min})\inf_{t \in [\tau_k,T^{*}]} f_t'(t)}{\alpha\theta_{\min} f_t(T^{*}) \Lambda(\tau_k,\Gamma(\tau_k))} (s_i - \Gamma_i(u)).
\end{align*}
Thus $\Gamma_i(u) \leq s_i - (s_i-\Gamma_i(\tau_k))\exp\left(\frac{(1-\alpha\theta_{\min}) \inf_{t \in [\tau_k,T^{*}]} f_t'(t)}{\alpha\theta_{\min} f_t(T^{*}) \Lambda(\tau_k,\Gamma(\tau_k))}(u-\tau_k)\right) < s_i$
for any time $u \in [\tau_k,T^{*})$.  As in the proof of Lemma~\ref{lemma:monotonic} we note that $\inf_{t \in [\tau,T^*]} f_t'(t)$ is attained as we are infimizing a continuous function over a compact space.  Thus, by continuity, this implies that $\Gamma_i(T^{*}) < s_i$ for all banks $i$.
\end{proof}

\subsection{Proof of Corollary~\ref{cor:n-unique}}
\begin{proof}
We will use Lemma~\ref{lemma:n-monotonic} to prove the existence and uniqueness of a solution. First, for all times $t \in [0,\tau_1]$ there exists a unique solution given by $\Gamma(t) = 0$, $q(t) = f_t(t)$, and $\Psi(t) = 0$.  
As in the Proof of Theorem~\ref{thm:unique}, we will consider the differential equation for $\Gamma$ given by \eqref{eq:dot-gamma2-n}.  We note that, though we considered the joint differential equation for $\Gamma$ and $q$ previously, \eqref{eq:dot-gamma2-n} only depends on $q$ through the collection of indicator functions on $\theta_i(t) \leq \theta_{\min}$; for the purposes of this proof we will replace the $i^\text{th}$ condition with $f_t(t)f_\Gamma(\sum_{j = 1}^n \Gamma_j(t)) \leq \bar q_i$.  From the solution $\Gamma$ we can immediately define $q(t) = f_t(t)f_\Gamma(\sum_{i = 1}^n \Gamma_i(t))$ for all times $t$, thus the existence and uniqueness of $\Gamma$ provides the same results for $q$.  The results for $\Psi$ follow from the same logic as $\Gamma$ and thus will be omitted herein.
We will consider an inductive argument to prove the existence and uniqueness.  Assume that we have the existence and uniqueness of the solution $\Gamma(t)$ up to time $\tau_k$ for some $k \in \{1,2,...,n\}$, then we wish to show we can continue this solution until $\tau_{k+1} \in [\tau_k,T]$.

By Lemma~\ref{lemma:n-monotonic}, $\dot \Gamma_i(\tau_k) \geq 0$ for all banks $i$. Define the process $\Gamma^*(t) = \sum_{i = 1}^n \Gamma_i(t) = \sum_{i = 1}^k \Gamma_i(t)$ with initial condition $\Gamma^*(\tau_k) = \sum_{i = 1}^{k-1} \Gamma_i(\tau_k)$.  Following the initial formulation for $\dot\Gamma_i(t)$ we find
\begin{align*}
\dot\Gamma^*(t) &= -\frac{Z_k^*(t,\Gamma^*(t))f_t'(t)f_\Gamma(\Gamma^*(t))}{1 + Z_k^*(t,\Gamma^*(t))f_t(t)f_\Gamma'(\Gamma^*(t))} \quad \text{with} \quad Z_k^*(t,\Gamma^*) = \frac{(1-\alpha\theta_{\min})[\sum_{i = 1}^k s_i - \Gamma^*]}{\alpha\theta_{\min}f_t(t)f_\Gamma(\Gamma^*)}\ind_{\{t \geq \tau_k\}}.
\end{align*}
We note that this follows the differential equation of the 1 bank setting (with possibly non-zero initial value).  Therefore we can conclude that $\Gamma^*(t)$ exists and is unique for $t \in [\tau_k,\tau_{k+1}]$ (where $\tau_{k+1}$ is a stopping time determined solely by $\Gamma^*$) via an application of Theorem~\ref{thm:unique}.
Utilizing this unique process $\Gamma^*$ we find that for any bank $i = 1,...,k$:
\[\dot\Gamma_i(t) = g_i(t,\Gamma) = \frac{(1-\alpha\theta_{\min})[f_t'(t)f_\Gamma(\Gamma^*(t)) + \dot\Gamma^*(t)f_t(t)f_\Gamma'(\Gamma^*(t))]}{\alpha\theta_{\min}f_t(t)f_\Gamma(\Gamma^*(t))}[s_i - \Gamma_i(t)].\] 
As $\Gamma^*(t)$ and $\dot\Gamma^*(t)$ are bounded in finite time we are able to deduce that $g_i$ is uniformly Lipschitz in $\Gamma$ and thus the existence and uniqueness of $\Gamma_i$ is guaranteed on the domain $[\tau_k,\tau_{k+1}]$.
\end{proof}

\section{Proofs from Section~\ref{sec:m-asset}}
\subsection{Proof of Lemma~\ref{lemma:nm-monotonic}}
\begin{proof}
We will consider this argument by induction.  
First, using Sylvester's determinant identity, we note that
$\operatorname{det}\left[I + Z(t,\Gamma) \diag[f_t(t)]\diag[f_\Gamma'(\Gamma^\T \vec{1})] s^\T\right] = \operatorname{det}\left[I + \diag[f_t(t)]\diag[f_\Gamma'(\Gamma^\T \vec{1})] s^\T Z(t,\Gamma)\right]$
for any $\Gamma \in \bbr^{n \times m}$.  Therefore, $\dot\Pi$ is well defined if and only if $\dot q$ is well defined.

Define $Y(t,\Gamma) := -Z(t,\Gamma) \diag[f_t(t)]\diag[f_\Gamma'(\Gamma^\T \vec{1})] s^\T$ and $r(t,\Gamma) = s \diag[\alpha \theta_{\min}] \diag[f_t(t)] f_\Gamma(\Gamma)$.  Note that $Y_{ij}(t,\Gamma) \geq 0$ and $r_i > 0$ for every $i,j = 1,...,n$ and all times $t$ and liquidation matrices $\Gamma$.  Therefore, utilizing results on the Leontief inverse, 
$I - Y(t,\Gamma)$ is invertible if $r(t,\Gamma)^\T Y(t,\Gamma) < r(t,\Gamma)^\T$.  The $i^\text{th}$ element of $r(t,\Gamma)^\T [I - Y(t,\Gamma)]$ can be expanded as:
\begin{align*}
r_i(t,\Gamma) &- \left[r(t,\Gamma)^\T Y(t,\Gamma)\right]_i\\
&= \sum_{k = 1}^m \left[\alpha_k \theta_{\min} f_{\Gamma,k}(\sum_{j = 1}^n \Gamma_{jk}) + (1-\alpha_k \theta_{\min}) f_{\Gamma,k}'(\sum_{j = 1}^n \Gamma_{jk}) \sum_{j = 1}^n (s_{jk} - \Gamma_{jk}) \ind_{\{\theta_j(t,\Gamma) \leq \theta_{\min}\}}\right]s_{ik} f_{t,k}(t)
\end{align*}
where $\theta_j(t,\Gamma)$ is the capital adequacy ratio for firm $j$ at time $t$ given the liquidation matrix $\Gamma$.
This is strictly greater than $0$ for every firm $i$ if 
$\alpha_k \theta_{\min} f_{\Gamma,k}(\sum_{j = 1}^n \Gamma_{jk}) + (1-\alpha_k \theta_{\min}) f_{\Gamma,k}'(\sum_{j = 1}^n \Gamma_{jk}) \sum_{j = 1}^n (s_{jk} - \Gamma_{jk}) \ind_{\{\theta_j(t,\Gamma) \leq \theta_{\min}\}} > 0$
for every asset $k$.  In particular, along the path of a solution $\Pi(t)$, the inequality in asset $k$ is equivalent to
\begin{align*}
0 &< \Lambda_k(t) := 1 + \frac{1-\alpha_k\theta_{\min}}{\alpha_k\theta_{\min}}\frac{\left[\sum_{j = 1}^n s_{jk} \left(1-\Pi_j(t)\right)\ind_{\{\theta_j(t) \leq \theta_{\min}\}}\right]f_\Gamma'(\sum_{j = 1}^n s_{jk} \Pi_j(t))}{f_\Gamma(\sum_{j = 1}^n s_{jk} \Pi_j(t))}
\end{align*}

For simplicity of notation, given a solution $\Pi(t)$, we will assume that the banks are ordered with increasing regulatory hitting times, i.e., $\tau_i \leq \tau_{i+1}$ for every firm $i$ (with the construction that $\tau_0 = 0$ and $\tau_{n+1} = T$) where $\tau_j := \inf\{t \in [0,T] \; | \; \theta_j(t) \leq \theta_{\min}\}$.
We will consider this proof by induction.  By $\Pi_i(t) = 0$ if $t < \tau_i$, the invertibility of $I - Y(t,\diag[\Pi(t)]s)$ is trivial for $t \in [0,\tau_1)$.  Now take $i \in \{1,2,...,n\}$ and $k \in \{1,2,...,m\}$.
For any time $t \in [\tau_i,\tau_{i+1})$ (or $t \in [\tau_i,T]$ if $\tau_{i+1} \geq T$)
\begin{align*}
\dot\Lambda_k(t) &= \frac{1-\alpha\theta_{\min}}{\alpha\theta_{\min}}\frac{d}{dt}\frac{\left[\sum_{j = 1}^i s_{jk}\left(1-\Pi_j(t)\right)\right]f_{\Gamma,k}'(\sum_{j = 1}^i s_{jk}\Pi_j(t))}{f_{\Gamma,k}(\sum_{j = 1}^i s_{jk}\Pi_j(t))}\\
&= \frac{1-\alpha\theta_{\min}}{\alpha\theta_{\min}}\frac{d}{d\Gamma_k}\left[\frac{\left[\left(\sum_{j = 1}^i s_{jk}\right)-\Gamma_k\right]f_{\Gamma,k}'(\Gamma_k)}{f_{\Gamma,k}(\Gamma_k)}\right]_{\Gamma_k = \sum_{j = 1}^i s_{jk}\Pi_j(t)} \sum_{j = 1}^i s_{jk}\dot\Pi_j(t).
\end{align*}
Using the same logic as in Lemma~\ref{lemma:monotonic}, we recover that $\dot\Lambda_k(t) \geq 0$ if and only if $\sum_{j = 1}^i s_{jk}\dot\Pi_j(t) \geq 0$ so long as 
$\frac{d}{d\Gamma_k}\left[\left(\left[\left(\sum_{j = 1}^i s_{jk}\right)-\Gamma_k\right]f_{\Gamma,k}'(\Gamma_k)\right)/f_{\Gamma,k}(\Gamma_k)\right] \geq 0$ at $\Gamma_k = \sum_{j = 1}^i s_{jk}\Pi_j(t)$.
This sufficient condition follows identically to the proof of Lemma~\ref{lemma:n-monotonic}.

Further, by construction, if $\Lambda_k(\tau_i) > 0$ for every asset $k$ then $\dot \Pi(\tau_i) \in \bbr^n_+$ by the non-negativity of the Leontief inverse and $\dot q(\tau_i) \in -\bbr^m_+$ by construction from $\dot \Pi(\tau_i)$.
We now want to demonstrate that $\Lambda_k(\tau_i) > 0$ for every asset $k$.  
Using the same construction as in the proof of Lemma~\ref{lemma:n-monotonic}, if $\alpha_k > -\frac{M_k f_{\Gamma,k}'(0)}{(1 - M_k f_{\Gamma,k}'(0))\theta_{\min}}$ then $\Lambda_k(\tau_i) > 0$.

By an argument identical to that in the proof of Lemma~\ref{lemma:monotonic}, we can show that $\Lambda_k(t) > 0$ for every asset $k$ and all times $t \in [\tau_i,\tau_{i+1})$ (or $t \in [\tau_i,T]$ if $\tau_{i+1} \geq T$).  Therefore by the above arguments and the non-negativity of the Leontief inverse, we can conclude that $\dot \Pi(t) \in \bbr^n_+$ (and thus $\dot q(t) \in -\bbr^m_+$) for all times $t \in [\tau_i,\tau_{i+1})$.

Finally, we will now demonstrate that $\Pi(t) \in [0,1)^n$ for all times $t \in [0,T]$.  We wish to consider two cases for this proof. If $\bar p_i \leq x_i + (1 - \alpha_{\ell,i} \theta_{\min}) \ell_i$ then bank $i$ will never be required to liquidate any assets (see Remark~\ref{rem:alphatheta}).  For the remainder of this proof we will consider the setting that $\bar p_i \geq x_i + (1 - \alpha_{\ell,i} \theta_{\min}) \ell_i$.  We wish to consider a decomposition of the capital ratio $\theta_i$ by $\theta_i(t) \geq \min_{k: \, s_{ik} > 0} \tilde\theta_{ik}(t)$ where:
\[\tilde\theta_{ik}(t) := \frac{c_{ik}x_i + \int_0^t s_{ik}\dot\Pi_i(u)q_k(u)du + s_{ik}(1 - \Pi_i(t))q_k(t) + c_{ik}\ell_i - c_{ik}\bar p_i}{\alpha_k s_{ik}(1 - \Pi_i(t))q_k(t) + c_{ik} \alpha_{\ell,i} \ell_i}\]
for all banks $i$ and assets $k$ (with $s_{ik} > 0$) where $\sum_{k = 1}^m c_{ik} = 1$.  In particular, we will choose the levels $c_{ik} \geq 0$ to be
$c_{ik} = \frac{(1 - \alpha_k \theta_{\min})s_{ik}}{\sum_{l = 1}^m (1 - \alpha_l \theta_{\min})s_{il}}$ for every bank $i$ and asset $k$.
This choice is made since $\tilde\theta_{ik}(0) \geq \theta_{\min}$ if and only if $c_{ik} \leq \frac{(1 - \alpha_k \theta_{\min})s_{ik}}{\bar p_i - x_i - (1 - \alpha_{\ell,i} \theta_{\min}) \ell_i} =: \bar c_{ik}$. It can trivially be shown that $\sum_{k = 1}^m \bar c_{ik} \geq 1$ if and only if $\theta_i(0) \geq \theta_{\min}$ which holds by assumption.  Therefore $c_{ik} = \bar c_{ik} / \sum_{l = 1}^m \bar c_{il} \leq \bar c_{ik}$ constructs a single asset problem with capital ratio $\tilde\theta_{ik}$ satisfying all the conditions of Lemma~\ref{lemma:n-monotonic} and Corollary~\ref{cor:n-unique} that can be solved independently.   Denote $\tilde\Gamma_{ik}$ to be the liquidation function for bank $i$ in asset $k$ so that the capital ratio $\tilde\theta_{ik}(t) \geq \theta_{\min}$ for all times $t$.  By Lemma~\ref{lemma:n-monotonic} it follows that $\tilde\Gamma_{ik}(t) < s_{ik}$.
By construction if $\tilde\theta_{ik}(t) \geq \theta_{\min}$ for every asset $k$ then $\theta_i(t) \geq \theta_{\min}$; thus setting $\tilde\Pi_i(t) := \max_{k = 1,...,m} \left\{\tilde\Gamma_{ik}(t)/s_{ik} \; | \; s_{ik} > 0\right\} < 1$ will guarantee $\Pi_i(t) \leq \tilde\Pi_i(t)$ for all times $t$ since selling more assets than necessary will drive the capital ratio above the regulatory threshold.
\end{proof}

\subsection{Proof of Corollary~\ref{cor:nm-unique}}
\begin{proof}
We will use Lemma~\ref{lemma:nm-monotonic} to prove the existence and uniqueness of a solution. First, for all times $t \in [0,\tau_1]$ there exists a unique solution given by $\Pi(t) = 0$, $q(t) = f_t(t)$, and $\Psi(t) = 0$.  
Following the same argument as in the proof of Corollary~\ref{cor:n-unique}, we will follow an inductive argument to prove the existence and uniqueness. 
For the purposes of this proof we will replace the $i^\text{th}$ condition $\theta_i(t) \geq \theta_{\min}$ with $\sum_{k = 1}^m (1-\alpha_k\theta_{\min})s_{ik}f_t(t)f_\Gamma(\sum_{j = 1}^n s_{jk} \Pi_j(t)) \leq \bar p_i - x_i - (1-\alpha_{\ell,i}\theta_{\min})\ell_i$ as, by construction, once a bank has hit the regulatory threshold it will remain there until the terminal time $T$.  
Assume that we have the existence and uniqueness of the solution $\Pi(t)$ up to time $\tau_k$ for some $k \in \{1,2,...,n\}$, then we wish to show we can continue this solution until $\tau_{k+1} \in [\tau_k,T]$.  As in the proof of Lemma~\ref{lemma:nm-monotonic}, we will reorder the banks so that between times $\tau_k$ and $\tau_{k+1}$, only the first $k$ banks have begun liquidating assets.  That is, $\theta_i(\tau_k) \leq \theta_{\min}$ if and only if $i \leq k$.

By Lemma~\ref{lemma:nm-monotonic}, $\dot \Pi_i(\tau_k) \geq 0$ for all banks $i$. 
Let $\Lambda$ be as in the proof of Lemma~\ref{lemma:nm-monotonic}, define the domain $U^k = \bigcap_{l = 1}^m \left\{\Pi \in [0,1]^n \; | \; \bar\Lambda_l^k(s^\T \Pi) \geq \frac{1}{2}\Lambda_l(\tau_k)\right\}$
where $\bar\Lambda^k: [0,M] \to \bbr^m$ is defined by
$\bar\Lambda_l^k(\Gamma) := 1 + \frac{1 - \alpha_l \theta_{\min}}{\alpha_l \theta_{\min}} \frac{\left[\left(\sum_{j = 1}^k s_{jl}\right) - \Gamma_l\right] f_{\Gamma,l}'(\Gamma_l)}{f_{\Gamma,l}(\Gamma_l)}$ for all $l = 1,...,m$.
We wish to note that from the above that by $\Lambda_l^k$ nondecreasing in time over $[\tau_k,\tau_{k+1}]$, any solution must lie in $U^k$, i.e., it must satisfy $\Pi(t) \in U^k$ for any time $t \in [\tau_k,\tau_{k+1}]$.  Thus by the logic of Theorem~\ref{thm:unique} we are able to conclude that there exists a unique solution on either $[\tau_k,\tau_{k+1}]$ or $[\tau_k,T^*)$ with $T^* \leq \tau_{k+1}$.  In the former case, the result is proven.  The latter is contradicted using the same bounding argument as in Theorem~\ref{thm:unique} with upper bound to $\dot\Pi$ provided in Lemma~\ref{lemma:nm-monotonic}.
\end{proof}

\section{Proofs from Section~\ref{sec:stresstest}}
For the following proofs, we will focus solely on the $m = 1$ asset framework with banks ordered by decreasing $\bar q$.  The general case is a result of the bounding argument in the proof of Lemma~\ref{lemma:nm-monotonic}.

\subsection{Proof of Theorem~\ref{thm:bound}}
\begin{proof}
We will prove this inductively for the single asset case.  Recall the definition of $\Lambda$ from the proof of Lemma~\ref{lemma:n-monotonic}, i.e., $\Lambda(t,\Gamma) = 1 + \sum_{j = 1}^n Z_j(t,\Gamma) f_t(t) f_\Gamma'(\sum_{j = 1}^n \Gamma_j)$.
\begin{enumerate}
\item First, by definition it is clear that $\tilde\tau_1 = \tau_1$ and $\tilde\Gamma(t) = \Gamma(t) = 0$ for all times $t \in [0,\tilde\tau_1]$.  Thus $\tilde\Lambda_1 = \Lambda(\tau_1,0)$ as well.  By the proof of Lemma~\ref{lemma:monotonic}, we know
$\dot\Gamma_1(t) \leq -\frac{(1-\alpha\theta_{\min}) f_t'(t)}{\alpha\theta_{\min} f_t(t) \tilde\Lambda_1} (s_1 - \Gamma_1(t))$
for $t \in [\tau_1,\tau_2]$.  As expressed in the proof of Lemma~\ref{lemma:monotonic} we can conclude $\tilde\Gamma_1^k(t) \geq \Gamma_1(t)$ for all times $t \in [\tilde\tau_1,\tau_2]$ and for any iteration $k = 1,...,n$ by construction as $\tilde\Gamma_1^1$ is the maximal solution to this differential inequality.
\item Fix $k \in \{1,2,...,n-1\}$ and assume $\tilde\Gamma_i^k(t) \geq \Gamma_i(t)$ for all times $t \in [0,\tau_{k+1}]$ and any firm $i = 1,...,k$.  This implies, for any $\hat k \geq k$, $\tilde\Gamma_i^{\hat k}(t) \geq \Gamma_i(t)$ for all times $t \in [0,\tau_{k+1}]$ as well.  Assume $\tau_{k+1} < T$ or else the proof is complete.  By monotonicity of the inverse demand function, $\tilde\tau_{k+1} \leq \tau_{k+1}$ with $\tilde\Gamma_i^k(\tilde\tau_{k+1}) \geq \Gamma_i(\tilde\tau_{k+1})$ for any $i = 1,...,k$.  In particular, this implies $\tilde\Lambda_{k+1} \geq \Lambda(\tilde\tau_{k+1},\Gamma(\tilde\tau_{k+1}))$.
By the proof of Lemma~\ref{lemma:n-monotonic} we can show
$\dot \Gamma_i(t) \leq -\frac{(1-\alpha\theta_{\min}) f_t'(t)}{\alpha\theta_{\min} f_t(t) \tilde\Lambda_{k+1}} (s_i - \Gamma_i(t))$
for $t \in [\tilde\tau_{k+1},\tau_{k+2})$ and firm $i = 1,...,k+1$.  We note that this is a stricter bound than that given in Lemma~\ref{lemma:n-monotonic}, but exists using the same logic.  Solving for the maximal solution to this differential inequality provides the solution $\tilde\Gamma^{k+1}$ which must satisfy $\tilde\Gamma_i^{k+1}(t) \geq \Gamma_i(t)$ for all times $t \in [0,\tau_{k+2},T]$.
\end{enumerate}
\end{proof}

\subsection{Proof of Corollary~\ref{cor:bound-exp}}
\begin{proof}
First, we will demonstrate that $\tilde\Gamma_i^n(t)$ has the expanded form provided.
\begin{align*}
\tilde\Gamma_i^n(t) &= \ind_{\{t < \tilde\tau_n\}} \tilde\Gamma_i^{n-1}(t) + \ind_{\{t \geq \tilde\tau_n\}} \left[s_i\left(1 - \left(\frac{f_t(t)}{f_t(\tilde\tau_n)}\right)^{\frac{1-\alpha\theta_{\min}}{\alpha\theta_{\min}\tilde\Lambda_n}}\right)  + \tilde\Gamma_i^{n-1}(\tilde\tau_n)\left(\frac{f_t(t)}{f_t(\tilde\tau_n)}\right)^{\frac{1-\alpha\theta_{\min}}{\alpha\theta_{\min}\tilde\Lambda_n}}\right]\\
&= \ind_{\{t < \tilde\tau_{n-1}\}} \tilde\Gamma_i^{n-2}(t) + \ind_{\{t \geq \tilde\tau_{n-1}\}} \tilde\Gamma_i^{n-1}(\tilde\tau_{n-1})\prod_{k = n-1}^n \left(\frac{f_t(t \wedge \tilde\tau_{k+1})}{f_t(t \wedge \tilde\tau_k)}\right)^{\frac{1-\alpha\theta_{\min}}{\alpha\theta_{\min}\tilde\Lambda_k}}\\
&\quad + \ind_{\{t \geq \tilde\tau_{n-1}\}} s_i \sum_{j = n-1}^n \left(1 - \left(\frac{f_t(t \wedge \tilde\tau_{j+1})}{f_t(t \wedge \tilde\tau_j)}\right)^{\frac{1-\alpha\theta_{\min}}{\alpha\theta_{\min}\tilde\Lambda_j}}\right)\prod_{k = j+1}^n \left(\frac{f_t(t \wedge \tilde\tau_{k+1})}{f_t(t \wedge \tilde\tau_k)}\right)^{\frac{1-\alpha\theta_{\min}}{\alpha\theta_{\min}\tilde\Lambda_k}}\\
&= \ind_{\{t < \tilde\tau_i\}} \tilde\Gamma_i^{i-1}(t) + \ind_{\{t \geq \tilde\tau_i\}} s_i \sum_{j = i}^n \left(1 - \left(\frac{f_t(t \wedge \tilde\tau_{j+1})}{f_t(t \wedge \tilde\tau_j)}\right)^{\frac{1-\alpha\theta_{\min}}{\alpha\theta_{\min}\tilde\Lambda_j}}\right)\prod_{k = j+1}^n \left(\frac{f_t(t \wedge \tilde\tau_{k+1})}{f_t(t \wedge \tilde\tau_k)}\right)^{\frac{1-\alpha\theta_{\min}}{\alpha\theta_{\min}\tilde\Lambda_k}}\\
&= s_i \sum_{j = i}^n \left(1 - \left(\frac{f_t(t \wedge \tilde\tau_{j+1})}{f_t(t \wedge \tilde\tau_j)}\right)^{\frac{1-\alpha\theta_{\min}}{\alpha\theta_{\min}\tilde\Lambda_j}}\right)\prod_{k = j+1}^n \left(\frac{f_t(t \wedge \tilde\tau_{k+1})}{f_t(t \wedge \tilde\tau_k)}\right)^{\frac{1-\alpha\theta_{\min}}{\alpha\theta_{\min}\tilde\Lambda_k}}\\
&= s_i\left(1 - \prod_{j = i}^n \left(\frac{f_t(t \wedge \tilde\tau_{j+1})}{f_t(t \wedge \tilde\tau_j)}\right)^{\frac{1-\alpha\theta_{\min}}{\alpha\theta_{\min}\tilde\Lambda_j}}\right).
\end{align*}
The penultimate line uses the fact that $\tilde\Gamma_i^{i-1}(t) = 0$ for all times $t$ by construction and $f_t(t \wedge \tilde\tau_{j+1})/f_t(t \wedge \tilde\tau_j) = 1$ for every $j \geq i$ if $t < \tilde\tau_i$.  

Now, let us consider the form of $\tilde\Lambda_i$ taking advantage of the exponential form for $f_\Gamma$:
\begin{align*}
\tilde\Lambda_i &= 1 + \frac{1-\alpha\theta_{\min}}{\alpha\theta_{\min}}\left[\sum_{j = 1}^{i}(s_j - \tilde\Gamma_j^{i-1}(\tilde\tau_i))\right]\frac{f_\Gamma'\left(\sum_{j = 1}^{i-1} \tilde\Gamma_j^{i-1}(\tilde\tau_i)\right)}{f_\Gamma\left(\sum_{j = 1}^{i-1} \tilde\Gamma_j^{i-1}(\tilde\tau_i)\right)}\\
&= 1 - b\frac{1-\alpha\theta_{\min}}{\alpha\theta_{\min}}\left[\sum_{j = 1}^i\left(s_j\left(\frac{f_t(\tilde\tau_i \wedge \tilde\tau_{j+1})}{f_t(\tilde\tau_i \wedge \tilde\tau_j)}\right)^{\frac{1-\alpha\theta_{\min}}{\alpha\theta_{\min}\tilde\Lambda_j}}\prod_{k = j+1}^n\left(\frac{f_t(\tilde\tau_i \wedge \tilde\tau_{k+1})}{f_t(\tilde\tau_i \wedge \tilde\tau_k)}\right)^{\frac{1-\alpha\theta_{\min}}{\alpha\theta_{\min}\tilde\Lambda_k}}\right)\right]\\
&= 1 - b\frac{1-\alpha\theta_{\min}}{\alpha\theta_{\min}}\left[\sum_{j = 1}^i s_j \prod_{k = j}^{i-1}\left(\frac{f_t(\tilde\tau_{k+1})}{f_t(\tilde\tau_k)}\right)^{\frac{1-\alpha\theta_{\min}}{\alpha\theta_{\min}\tilde\Lambda_k}}\right]\\
\end{align*}

Finally, let us consider the time at which the analytical worst-case pricing process hits $\bar q_i$, i.e., the time when firm $i$ reaches the regulatory threshold $\theta_{\min}$ provided all firms follow the worst-case path.  As no firms act before $\tilde\tau_1 = \tau_1$, this can easily be computed as $\tilde\tau_1 = f_t^{-1}(\bar q_1)$.  Consider now $i = 2,...,n$, recall that $\bar q_1 \geq \bar q_2 \geq ... \geq \bar q_n$, and assume $t \geq \tilde\tau_{i-1}$:
\begin{align*}
& \bar q_i = f_t(t)f_\Gamma\left(\sum_{j = 1}^{i-1} \tilde\Gamma_j^n(t)\right) 
\Leftrightarrow \bar q_i = f_t(t)f_\Gamma\left(\sum_{j = 1}^{i-1} s_j - \sum_{j = 1}^{i-1} s_j \prod_{k = j}^{i-1}\left(\frac{f_t(t \wedge \tilde\tau_{k+1})}{f_t(\tilde\tau_k)}\right)^{\frac{1-\alpha\theta_{\min}}{\alpha\theta_{\min}\tilde\Lambda_k}}\right)\\
&\Leftrightarrow \log(\bar q_i) - \sum_{j = 1}^{i-1} s_j = \log(f_t(t)) + b\sum_{j = 1}^{i-1} s_j \prod_{k = j}^{i-1}\left(\frac{f_t(t \wedge \tilde\tau_{k+1})}{f_t(\tilde\tau_k)}\right)^{\frac{1-\alpha\theta_{\min}}{\alpha\theta_{\min}\tilde\Lambda_k}}\\
&\Leftrightarrow \log(\bar q_i) - \sum_{j = 1}^{i-1} s_j = \log(f_t(t)) + \left(\frac{b f_t(t)^{\frac{1-\alpha\theta_{\min}}{\alpha\theta_{\min}\tilde\Lambda_{i-1}}}}{f_t(\tilde\tau_{i-1})^{\frac{1-\alpha\theta_{\min}}{\alpha\theta_{\min}\tilde\Lambda_{i-1}}}}\right)\left[\sum_{j = 1}^{i-1} s_j \prod_{k = j}^{i-2}\left(\frac{f_t(\tilde\tau_{k+1})}{f_t(\tilde\tau_k)}\right)^{\frac{1-\alpha\theta_{\min}}{\alpha\theta_{\min}\tilde\Lambda_k}}\right]\\
&\Leftrightarrow \log(\bar q_i) - \sum_{j = 1}^{i-1} s_j = \log(f_t(t)) + \left(\frac{\alpha\theta_{\min}}{1-\alpha\theta_{\min}}\right) \nu_{i-1} f_t(t)^{\frac{1-\alpha\theta_{\min}}{\alpha\theta_{\min}\tilde\Lambda_{i-1}}}\\
&\Leftrightarrow f_t(t) = \left[\frac{\tilde\Lambda_{i-1} W\left(\frac{\nu_{i-1}}{\tilde\Lambda_{i-1}}\exp\left(\frac{1-\alpha\theta_{\min}}{\alpha\theta_{\min}\tilde\Lambda_{i-1}}\left[\log\left(\bar q_i\right) + b\sum_{j = 1}^{i-1}s_j\right]\right)\right)}{\nu_{i-1}}\right]^{\frac{\alpha\theta_{\min}\tilde\Lambda_{i-1}}{1-\alpha\theta_{\min}}}
\end{align*}
\end{proof}

\subsection{Proof of Corollary~\ref{cor:probability}}
\begin{proof}
First, before we prove the bound provided in Corollary~\ref{cor:probability} we need to demonstrate that $\tilde\Lambda_k$ and $\nu_k$ do not depend on the parameter $a$ of the inverse demand function $f_t$, i.e., they are constants in this problem.  We will do this by induction jointly on $\tilde\Lambda_k$, $\nu_k$, and $f(\tilde\tau_k)$ for $k = 1,...,n$ (trivially this is the case for the assumed values $\tilde\Lambda_0 = 1$, $\nu_0 = 0$, and $f(\tilde\tau_0) = 1$).
\begin{enumerate}
\item Fix $k = 1$, then $\tilde\Lambda_1 = 1 - b\frac{1-\alpha\theta_{\min}}{\alpha\theta_{\min}}s_1$, $f(\tilde\tau_1) = \bar q_1$, and $\nu_1 = \frac{1-\tilde\Lambda_1}{f_t(\tilde\tau_1)^{\frac{1-\alpha\theta_{\min}}{\alpha\theta_{\min}\tilde\Lambda_1}}}$.  Since $\tilde\Lambda_1$ and $f(\tilde\tau_1)$ do not depend on the parameter $a$ then neither does $\nu_1$.
\item Fix $k \in \{2,...,n\}$ and assume $(\tilde\Lambda_i,\nu_i,f(\tilde\tau_i))_{i = 1}^{k-1}$ do not depend on the parameter $a$.  By Corollary~\ref{cor:bound-exp}, $f(\tau_k)$ only depends on $\tilde\Lambda_{k-1}$ and $\nu_{k-1}$, and thus does not depend on the parameter $a$.  Additionally, $\tilde\Lambda_k$ only depends on $(f_t(\tilde\tau_i))_{i = 1}^k$, which (from the prior statement) does not depend on $a$.  Finally, $\nu_k$ only depends on $(\tilde\Lambda_i,f_t(\tilde\tau_i))_{i = 1}^k$, thus it does not depend on $a$ either.
\end{enumerate}

We will prove the bound on the probability by induction:
\begin{enumerate}
\item Let $q^* \in [\bar q_1,1]$ (i.e., $k = 0$).  For such an event to occur, no firms will have hit the regulatory threshold and thus it must be the case that $q(t) = f_t(t)$.  Therefore,
\begin{align*}
\P(q(t) \geq q^*) &= \P(f_t(t) \geq q^*) = \P\left(a \leq -\frac{1}{t}\log(q^*)\right) = \P\left(a \leq \frac{1}{t}\Phi_0^{-1}(\log(q^*))\right).
\end{align*}
\item Assume the provided bound is true for any $q^* \in [\bar q_k,1]$.  Now let $q^* \in [\bar q_{k+1},\bar q_k)$.
\begin{align*}
\P(q(t) \geq q^*) &= \P(q(t) \geq \bar q_k) + \P(q(t) \in [q^*,\bar q_k))\\
&\geq \P\left(a \leq \Phi_{k-1}^{-1}(\log(q^*) + b\sum_{i = 1}^{k-1} s_i)\right) + \P\left(f_t(t)f_\Gamma\left(\sum_{i = 1}^k \tilde\Gamma_i^n(t)\right) \in [q^*,\bar q_k)\right)
\end{align*}
Now we wish to show the form for the last term in our bound.
\begin{align*}
\P&\left(f_t(t)f_\Gamma\left(\sum_{i = 1}^k \tilde\Gamma_i^n(t)\right) \in [q^*,\bar q_k)\right) = \P\left(-at - b\sum_{i = 1}^k \tilde\Gamma_i^n(t) \in [\log(q^*),\log(\bar q_k))\right)\\
&= \P\left(-at + \left(\frac{\alpha\theta_{\min}}{1-\alpha\theta_{\min}}\right) \nu_k f_t(t)^{\frac{1-\alpha\theta_{\min}}{\alpha\theta_{\min}\tilde\Lambda_k}}
 \in [\log(q^*) + b\sum_{i = 1}^k s_i , \log(\bar q_k) + b\sum_{i = 1}^k s_i)\right)\\
&= \P\left(-at + \left(\frac{\alpha\theta_{\min}}{1-\alpha\theta_{\min}}\right) \nu_k \exp\left(-at\frac{1-\alpha\theta_{\min}}{\alpha\theta_{\min}\tilde\Lambda_k}\right)
 \in [\log(q^*) + b\sum_{i = 1}^k s_i , \log(\bar q_k) + b\sum_{i = 1}^k s_i)\right)\\
&= \P\left(a \in (\Phi_k^{-1}\left(\log(\bar q_k) + b\sum_{i = 1}^k s_i\right) , \Phi_k^{-1}\left(\log(q^*) + b\sum_{i = 1}^k s_i\right)]\right).
\end{align*}
The result follows from $\Phi_{k-1}^{-1}(\log(\bar q_k) + b\sum_{i = 1}^{k-1} s_i) = \Phi_k^{-1}(\log(\bar q_k) + \sum_{i = 1}^k s_i)$ as shown below:
\begin{align*}
&\Phi_{k-1}^{-1}\left(\log(\bar q_k) + b\sum_{i = 1}^{k-1} s_i\right) = \left(\frac{\alpha\theta_{\min}}{1-\alpha\theta_{\min}}\right) \nu_{k-1} f_t(\tilde\tau_k)^{\frac{1-\alpha\theta_{\min}}{\alpha\theta_{\min}\tilde\Lambda_{k-1}}} - \left[\log(\bar q_k) + b\sum_{i = 1}^{k-1} s_i\right]\\
&= \left(\frac{\alpha\theta_{\min}}{1-\alpha\theta_{\min}}\right) \nu_k f_t(\tilde\tau_k)^{\frac{1-\alpha\theta_{\min}}{\alpha\theta_{\min}\tilde\Lambda_k}} - \left[\log(\bar q_k) + b\sum_{i = 1}^k s_i\right]\\
&= \left(\frac{\alpha\theta_{\min}\tilde\Lambda_k}{1-\alpha\theta_{\min}}\right)\left(\frac{1-\tilde\Lambda_k}{\tilde\Lambda_k}\right) - \left[\log(\bar q_k) + b\sum_{i = 1}^k s_i\right]\\
&= \left(\frac{\alpha\theta_{\min}\tilde\Lambda_k}{1-\alpha\theta_{\min}}\right)W\left(\frac{1-\tilde\Lambda_k}{\tilde\Lambda_k}\exp\left[\frac{1-\tilde\Lambda_k}{\tilde\Lambda_k}\right]\right) - \left[\log(\bar q_k) + b\sum_{i = 1}^k s_i\right]\\
&= \left(\frac{\alpha\theta_{\min}\tilde\Lambda_k}{1-\alpha\theta_{\min}}\right)W\left(\left(\frac{\nu_k}{\tilde\Lambda_k}\right)f_t(\tilde\tau_k)^{\frac{1-\alpha\theta_{\min}}{\alpha\theta_{\min}\tilde\Lambda_k}}\exp\left[\frac{1-\alpha\theta_{\min}}{\alpha\theta_{\min}\tilde\Lambda_k}b\sum_{i = 1}^k[s_i - \tilde\Gamma_i^n(\tilde\tau_k)]\right]\right) - \left[\log(\bar q_k) + b\sum_{i = 1}^k s_i\right]\\
&= \Phi_k^{-1}\left(\log(\bar q_k) + b\sum_{i = 1}^k s_i\right).
\end{align*}
\end{enumerate}
\end{proof}

\bibliographystyle{plain}
\bibliography{bibtex2}

\end{document}